\documentclass{jfm}
\usepackage{graphicx}
\usepackage{epstopdf, epsfig}
\usepackage{siunitx}

\usepackage[finalnew]{trackchanges}
\addeditor{Yu}

\shorttitle{Turbulence Spectra in the Stable Atmospheric Boundary Layer}
\shortauthor{Yu Cheng, Qi Li, Stefania Argentini, Chadi Sayde, Pierre Gentine}

\title{Turbulence Spectra in the Stable Atmospheric Boundary Layer}

\author{Yu Cheng\aff{1}
  \corresp{\email{yc2965@columbia.edu}},
  Qi Li\aff{1,2},
  Stefania Argentini\aff{3},
  Chadi Sayde\aff{4} \\
   \and Pierre Gentine\aff{1}}

\affiliation{\aff{1}Department of Earth and Environmental Engineering, Columbia University, New York, \\ NY 10027, USA
\aff{2}Department of Civil and Environmental Engineering, Cornell University, Ithaca, \\ NY 14853, USA\aff{3}Institute of Atmospheric Sciences and Climate, CNR, Rome, Italy\aff{4}Department of Biological and Agricultural Engineering, North Carolina State University, Raleigh, NC 27695, USA}

\begin{document}

\maketitle

\begin{abstract}
Stratification can cause turbulence spectra to deviate from Kolmogorov's isotropic -5/3 power-law scaling in the universal equilibrium range at high Reynolds numbers. However, a consensus has not been reached with regard to the exact shape of the spectra. Here we propose a theoretically-derived shape of the \add{turbulent kinetic energy} ($\mathrm{TKE}$) and temperature spectra \add{in horizontal wavenumber} that consists of three regimes at small Froude number: the buoyancy subrange, a transition region and isotropic inertial subrange through derivation based on previous research. These regimes are confirmed by various observations in the atmospheric boundary layer. We also show that $\mathrm{DNS}$ may not apply in the study of very stable atmospheric boundary layers \add{at very high Reynolds numbers} as they cannot correctly represent the observed spectral regimes because of the lack of scale separation \add{limited by current computational capacity}. In addition, the spectrum in the transition regime explains why Monin-Obukhov similarity theory cannot entirely describe the behavior of the stable atmospheric boundary.
\end{abstract}

\section{Introduction}
Isotropic turbulence spectra follow a -5/3 power-law scaling in the inertial subrange \citep{kolmogorov1941}. However, stratification \citep{lilly1983stratified} is widely observed in geophysical turbulence, for example, in nocturnal atmospheric boundary layer \citep{mahrt1998stratified}, in most part of troposphere \citep{fiedler1970atmospheric}, above the tropopause  \citep{nastrom1985climatology,tulloch2006theory} and below ocean mixed layer \citep{baker1987sampling} and generates anisotropy. Because of this stratification \citep{bolgiano1959turbulent,dougherty1961anisotropy,lumley1964spectrum,phillips1965bolgiano,weinstock1978theory}, the spectra deviate from the -5/3 scaling of isotropic turbulence. At very large, synoptic scales, theory and observation have shown that horizontal wavenumber spectra of turbulence kinetic energy ($\mathrm{TKE}$) exhibit a -3 power-law scaling \citep{charney1971geostrophic,nastrom1985climatology}. In the mesoscale regime, a scaling close to -5/3 is still observed \citep{nastrom1985climatology}. Recently, a direct energy cascade hypothesis \citep{lindborg2006energy} was supported by \change{Direct Numerical Simulation}{direct numerical simulation} ($\mathrm{DNS}$) studies \citep{brethouwer2007scaling,kimura2012energy,riley2003dynamics,waite2004stratified} to explain the -5/3 scaling of anisotropic turbulence.

In the atmospheric surface layer (at lower part of the boundary layer), and thus at high Reynolds numbers, it is widely accepted that Monin-Obukhov similarity theory (MOST) \citep{monin1954basic} works well in neutral or weakly stable cases but does not apply in very stable cases \citep{mahrt1998stratified,mahrt2014stably}. Several hypotheses have been proposed to account for the failure of MOST, such as upside-down boundary layers \citep{mahrt1985vertical,ohya1997turbulence}, mesoscale motions \citep{mahrt1999stratified,smeets1998turbulence}, surface heterogeneity \citep{derbyshire1995stable} and Kelvin-Helmholtz instability with discontinuous and intermittence turbulence \citep{cheng2005pathology}. When applying MOST to calculate velocity and temperature variance, the -5/3 spectra scaling in the equilibrium range is always assumed \citep{kaimal1972spectral}.

\remove{Within the universal equilibrium range of ``approximately isotropic" turbulence, Weinstock (1978) showed that stratification results in a transition region between a ``buoyancy subrange" and the ``isotropic inertial subrange", where the spectra do not follow a universal power law. Here in ``approximately isotropic" turbulence without effects of the wall, the �buoyancy subrange” denotes scales larger than buoyancy length scale (Billant \& Chomaz 2011) and ``isotropic inertial subrange" denotes scales smaller than the Dougherty-Ozmidov length scale
(Dougherty 1961;Ozmidov 1965), which are different than those defined in Weinstock (1978). This transition region is the same as the ``spectral bump'' region observed recently in numerical simulations (Khani \& Waite 2014;Waite 2011) near the buoyancy scale.}

\cite{weinstock1978theory} derived the stratified turbulence spectra by assuming ``approximately isotropic" fluid motions.  For eddies below ``energy containing" scales in the atmospheric boundary layer, we relax the ``approximately isotropic" hypothesis for horizontal wavenumber spectra, which consist of buoyancy subrange, transition region and isotropic inertial subrange.
We will show that the transition region is related to three scales: the buoyancy length scale \citep{billant2001self}, the Dougherty-Ozmidov scale \citep{dougherty1961anisotropy,ozmidov1965turbulent} and the distance from wall \citep{townsend1976structure,katul2014two} in \add{horizontal wavenumber spectra of} both $\mathrm{TKE}$ and temperature in the equilibrium range at high Reynolds numbers. We further suggest that the deviation from the -5/3 power-law scaling in the transition region leads to the failure in applying MOST to very stable atmospheric boundary layers \add{(ABL)}.
\section{Derivation of spectra scaling}\label{sec:Derivation of spectra scaling}
     \subsection{Scale definition}
%The equations of motion for incompressible fluid under the Boussinesq approximation are
%\begin{equation}
%\frac{\p \boldsymbol{u}}{\p t}+\boldsymbol{u} \cdot \nabla{\boldsymbol u}=-\nabla{p}+b'\boldsymbol{\hat{z}}+ \boldsymbol{F}_u+D_u(\boldsymbol{u}),
%\end{equation}
%\begin{equation}
%\nabla \cdot \boldsymbol u=0,
%\end{equation} 
% \begin{equation}
% \frac{\p b'}{\p t}+\boldsymbol{u} \cdot \nabla{b'}+N^2w=F_{b'}+D_{b'}(b'),
% \end{equation}
% where $\boldsymbol{u}=u\boldsymbol{\hat{x}}+v\boldsymbol{\hat{y}}+w\boldsymbol{\hat{z}}$ 
% is velocity, $b'=-g \rho' / \rho_0 $ or $ -g \theta' / \theta_0 $ is buoyancy, $g$ is gravity acceleration, $ \rho' $ and $\theta'$ are density and potential temperature perturbations, $ \rho_0 $ and $ \theta_0 $ are mean values, $p$ is dynamic pressure divided by $ \rho_0 $, $N$ is the Brunt-V\"{a}is\"{a}l\"{a} frequency, $ \boldsymbol{F}_u $ and $F_{b'}$ are forcing terms, and $D_u$ and $D_{b'}$ are dissipation terms. 
 
 The strength of the stratification is assessed using the horizontal Froude number \change{$Fr=\frac{U}{NL_h}$}{$Fr_h=\frac{U}{NL_h}$},
 %\begin{equation}
%Fr=\frac{U}{NL_h},
% \end{equation}
where $U$ is root mean square of the horizontal velocity, $N$ is the Brunt-V\"{a}is\"{a}l\"{a} frequency and $L_h$ is a horizontal length scale. The horizontal length scale $L_h=\frac{U^3}{\epsilon}$ is determined by invoking Taylor's frozen turbulence hypothesis \citep{taylor1938spectrum}, where $\epsilon$ is $\mathrm{TKE}$ dissipation rate. The Reynolds number is $Re=\frac{UL_h}{\nu}$, where $\nu$ is kinetic viscosity and the buoyancy Reynolds number is \change{$Re_b=ReFr^2$}{$Re_b=ReFr_h^2$} \citep{brethouwer2007scaling}. Buoyancy length scale is $L_b=2\upi{\frac{U}{N}}$ \citep{billant2001self} and the corresponding wavenumber is
\begin{equation}
 k_b=\frac{2\upi}{L_b}.
\end{equation}
The Dougherty-Ozmidov length scale is $L_O=2\upi\big( \frac{\epsilon}{N^3} \big)^{1/2}$ \citep{dougherty1961anisotropy,ozmidov1965turbulent}, where the stratification effect becomes important. The corresponding wavenumber to Dougherty-Ozmidov length scale is
\begin{equation}
k_O=\frac{2\upi}{L_O}. 
\end{equation} 
As Dougherty and Ozmidov independently defined the length scale, we call it Dougherty-Ozmidov length scale following \cite{grachev2015similarity}. We thus assume that isotropic turbulence applies only for length scales below $L_O$ when there is no wall. The ratio of the buoyancy scale to Dougherty-Ozmidov scale is related to the horizontal Froude number as \change{$\frac{L_b}{L_O}=Fr^{-1/2}$}{$\frac{L_b}{L_O}=Fr_h^{-1/2}$}.
%\begin{equation}
%\frac{L_b}{L_O}=Fr^{-1/2}.
%\end{equation}
Therefore, small \change{$Fr$ ($Fr\ll1$)}{$Fr_h$ ($Fr_h\ll1$)} is required for a scale separation between $L_b$ and $L_O$. The Kolmogorov length scale \citep{kolmogorov1941} is $\eta=\big( \nu^3/\epsilon \big)^{1/4} $ and the ratio of Dougherty-Ozmidov to Kolmogorov length scale is \change{$\frac{L_O}{\eta}=Re_b^{3/4}$}{$\frac{L_O}{\eta}=2\pi Re_b^{3/4}$}.
%\begin{equation}
%\frac{L_O}{\eta}=Re_b^{3/4}.
%\end{equation}
So large $Re_b$ ($Re_b\gg1$ \citep{brethouwer2007scaling}) is required for the separation between $L_O$ and $\eta$. In the rest of this manuscript we thus assume that both \change{$Fr\ll1$}{$Fr_h\ll1$} and $Re_b\gg1$ conditions are met.
\subsection{Spectra shape}   
We start from the spectral $\mathrm{TKE}$ balance equation \citep{lumley1964structure}
\begin{equation}
\frac{\p E(k)}{\p t}+\frac{\p Q(k)}{\p z}=S(k)\frac{\p U_0}{\p z}-\frac{\p \epsilon(k)}{\p k}+B(k)-2\nu k^2E(k),
\end{equation}  
where $E(k)$ is spectral kinetic energy density, $\frac{\p Q(k)}{\p z}$ is the vertical transfer of turbulent energy in physical space, $S(k)$ is spectrum of the Reynolds stress $-\overline{uw}$, $u$ and $w$ are streamwise and vertical velocity fluctuation respectively, $U_0$ is mean streamwise velocity, $\epsilon(k)$ is net rate of spectral energy transfer, $B(k)$ is spectrum of buoyancy flux $-\frac{g}{\rho_0}\overline{w\rho'} $ and $2\nu k^2E(k)$ is rate of energy dissipation by molecular viscosity $\nu$. Assuming steady state, neglecting $\frac{\p Q(k)}{\p z}$ for eddies smaller than the energy containing scales and neglecting molecular viscosity (i.e. for eddies much larger than Kolmogorov scale),  \cite{weinstock1978theory} obtained  $\frac{\p \epsilon(k)}{\p k}=S(k)\frac{\p U_0}{\p z}+B(k)$ for universal equilibrium range.
%\begin{equation}
%\frac{\p \epsilon(k)}{\p k}=S(k)\frac{\p U_0}{\p z}+B(k).
%\end{equation}
The buoyancy flux spectrum can also be written as $B(k)=-R_f S(k) \frac{\p U_0}{\p z}$ \citep{lumley1965theoretical},    
%\begin{equation}
%B(k)=-R_f S(k) \frac{\p U_0}{\p z}
%\end{equation}
where $R_f=\frac{g}{T_0} \frac{\overline{wT'}}{\overline{uw} \frac{\p U_0}{\p z}}$ is flux Richardson number and $T'$ is fluctuation from mean temperature $\overline{T}$. Substituting $B(k)$ into spectral balance equation, \cite{weinstock1978theory} obtained $\frac{\p \epsilon(k)}{\p k}=B(k)\Big( 1-\frac{1}{R_f} \Big),$
%\begin{equation}
%\frac{\p \epsilon(k)}{\p k}=B(k)\Bigg( 1-\frac{1}{R_f} \Bigg),
%\end{equation}
where $R_f$ is larger than $0$ for stratified turbulence. Weinstock did not realize that there is an upper bound \citep{townsend1958turbulent,nieuwstadt1984turbulent,schumann1995turbulent,zilitinkevich2013hierarchy} for $R_f$ under the steady state assumption. Here we will take the uper bound of $R_f$ to be $0.25$ \citep{nieuwstadt1984turbulent,zilitinkevich2013hierarchy,katul2014two} for the existence of continuous turbulence with Richardson-Kolmogorov cascade, which is discussed in detail in \cite{grachev2013critical}. Assuming approximately isotropic turbulence, \cite{weinstock1978theory} obtained 
%$B(k)=-\alpha a N^2 \epsilon(k)^{\frac{2}{3}} \frac{v_m k^{-2/3}}{0.8 N^2+k^2 v_m^2}$,
\begin{equation}
B(k)=-\alpha a N^2 \epsilon(k)^{\frac{2}{3}} \frac{v_m k^{-2/3}}{0.8 N^2+k^2 v_m^2}, \label{eq:2.40}
\end{equation}
where $\alpha=1.5$ is the Kolmogorov constant \citep{sreenivasan1995universality,pope2000turbulent}, $a$ is anisotropic factor (we will have detailed discussion for $a$) given as $1$ for isotropic turbulence and 0.5 for ``approximately isotropic" turbulence. $v_m$ is the root mean square of fluctuating velocity for eddies smaller than energy containing range given as $v_m^2=\frac{2}{3} \int_{k_m}^{\infty}E(k) \mathrm{d} k$,
%\begin{equation}
%v_m^2=\frac{2}{3} \int_{k_m}^{\infty}E(k) \mathrm{d} k.
%\end{equation}     
where $k_m$ is the smallest wavenumber in the universal equilibrium range. When $0.8N^2 \gg k^2 v_m^2$, the effect of buoyancy denoted by $N$ is large and $B(k)$ is in the buoyancy subrange; 
when $0.8N^2 \ll k^2 v_m^2$, the effect of buoyancy is small and $B(k)$ is in the inertial subrange. The transition wavenumber was defined \citep{weinstock1978theory} as $k_{BW}=\frac{0.8^{1/2} N}{v_m}$,
%\begin{equation}
%k_{BW}=\frac{0.8^{1/2} N}{v_m},
%\end{equation}
whose corresponding length scale is $ L_{BW}=\frac{2 \upi}{k_{BW}}$.  \add{Note that the above analysis by Weinstock assumes a similar spectra shape between horizontal wavenumber and vertical wavenumber spectra, i.e., ``approximately isotropic". However, that assumption does not apply in our hypothesis as vertical wavenumber spectra are not considered in the atmospheric boundary layer.} We will relax the ``approximately isotropic" turbulence hypothesis below \add{by only considering horizontal wavenumber spectra, e.g., equation (2.4) will be used only for horizontal wavenumber.}
 
We choose the buoyancy scale $L_b$ to be the minimum scale of buoyancy subrange, which requires $1.118 \frac{v_m}{U} \ll 1$. If the Dougherty-Ozmidov length scale $L_O$ is the maximum scale of isotropic inertial subrange for turbulence without walls, then the wavenumber $k_O=\frac{2 \upi}{L_O}$ will have to satisfy $0.8N^2 \ll k_O^2 v_m^2$, which requires \change{$1.118 \frac{v_m}{U} \gg Fr^{ \frac{1}{2}}$}{$1.118 \frac{v_m}{U} \gg Fr_h^{ \frac{1}{2}}$}. In order to observe the transition region between the Dougherty-Ozmidov scale $L_O$ and buoyancy scale $L_b$, we thus need to satisfy the condition:
\begin{equation}
Fr_h^{\frac{1}{2}} \ll 1.118 \frac{v_m}{U}  \ll 1.
\end{equation}
In the equilibrium range of \add{horizontal wavenumber} in stably stratified turbulence without walls, scales above $L_b$ will thus be in the buoyancy subrange, scales below $L_O$ will be in the isotropic inertial subrange, and scales in between will be in a transition region. 

However, like \cite{weinstock1978theory}, the above analysis does not consider the effects of wall on turbulence in the atmospheric boundary layer. Equation ($19a'$) in \cite{weinstock1978theory} assumes a similar spectra shape between $w$ and $u$ in horizontal wavenumber as the focus was ``approximately isotropic" turbulence without considering wall effects. In the atmospheric boundary layer, the existence of wall will cause $w$ spectrum to deviate from $u$ spectrum at low horizontal wavenumbers. Field observation \citep{grachev2013critical} and models \citep{katul2014two} suggest a shallower slope than -5/3 at low horizontal wavenumbers in $w$ spectrum. Following \cite{katul2014two}, we consider the wavenumber $k_a$ that satisfies 
\begin{equation}
k_az=1, 
\end{equation}
where $z$ is height above ground. 
If $k_a \le k_O$, we assume the aforementioned anisotropic factor $a$ to be as follows
\begin{equation}
a= \left\{
\begin{array}{ll}
1, & k \ge k_O, \\[2pt]
\big(\frac{k}{k_O} \big)^{2/3} , & k < k_O.
\end{array} \right.
\end{equation}
If $k_a>k_O$, we assume the anisotropic factor $a$ to be as follows
\begin{equation}
a= \left\{
\begin{array}{ll}
1, & k \ge k_a, \\[2pt]
\big(\frac{k}{k_a} \big)^{2/3} , & k < k_a.
\end{array} \right.
\end{equation}
Under this assumption, the $w$ spectrum will be different from $u$ spectrum when 
$k<\max (k_O,k_a)$ in horizontal wavenumber and $w$ variance will be smaller than $u$ variance as $k$ decreases. However, \change{We}{we} can still assume a -5/3 scaling for \add{horizontal wavenumber} TKE spectrum as the variance of horizontal velocity largely dominates at low wavenumbers. Note that we only consider horizontal wavenumber spectrum in the atmospheric boundary layer as turbulence in the vertical direction is not homogeneous \citep{fiedler1970atmospheric}. \add{Contrary to Weinstock's theory, we do not assume a similar shape between vertical wavenumber spectra and horizontal wavenumber spectra because we only consider horizontal wavenumber spectra in the ABL.} 
Calculations for buoyancy flux and energy spectrum will then differ from \cite{weinstock1978theory} when $k<\max (k_O,k_a)$. 
The energy spectrum in horizontal wavenumber can be obtained for the universal equilibrium range as \citep{weinstock1978theory}
\begin{equation}
E(k)=\alpha \epsilon_0^{\frac{2}{2}} {\Bigg[ {1-\frac{5\alpha^{\frac{2}{3}}}{12}} \frac{v_m}{v_0} \Big( \frac{1-R_f}{R_f}   \Big) C \Big( \frac{k}{k_{BW}} \Big) \Bigg]}^2 k^{-\frac{5}{3}},
\end{equation}
where $v_0^2=\frac{2}{3} \int_{k_{BW}}^{\infty} E(k) \mathrm{d} k $, $\epsilon_0$ is viscous dissipation rate %\citep{phillips1965bolgiano}
and $C\Big( \frac{k}{k_{BW}} \Big) $ in \cite{weinstock1978theory} satisfies
\begin{equation}
C\Big( \frac{k}{k_{BW}} \Big)= \left\{
\begin{array}{ll}
\frac{3-3 {\big( \frac{k}{k_{BW}} \big) }^{\frac{1}{3} }} {1+\frac{1}{5} {\big( \frac{k}{k_{BW}} \big) }^2}+0.5, & k < k_{BW}, \\[2pt]
\frac{\frac{3}{5} {\big( \frac{k}{k_{BW}} \big) }^{-\frac{5}{3} }} {1+\frac{1}{5} {\big( \frac{k}{k_{BW}} \big) }^{-2}}, & k \ge k_{BW}.
\end{array} \right.
\end{equation}
Details of the functional forms of $C\Big( \frac{k}{k_{BW}} \Big) $ under different cases in the ABL are discussed as follows.

\subsubsection{Case $k_a \le k_O$}   
If $k_a \le k_O$, similarly to \cite{weinstock1978theory}, we obtain 
\begin{equation}
C\Big( \frac{k}{k_{BW}} \Big)= \left\{
\begin{array}{ll}
\frac{\frac{3}{5} {\Big( \frac{k_O}{k_{BW}} \Big) }^{-\frac{5}{3} }} {1+\frac{1}{5} {\Big( \frac{k_O}{k_{BW}} \Big) }^{-2}}+ { \Big( \frac{k_{BW}}{k_O} \Big)}^{\frac{2}{3}} \Big[ \arctan \Big( \frac{k_O}{k_{BW}} \Big) - \arctan \Big( \frac{k}{k_{BW}} \Big)    \Big], & k < k_O, \\[4pt]
\frac{\frac{3}{5} {\Big( \frac{k}{k_{BW}} \Big) }^{-\frac{5}{3} }} {1+\frac{1}{5} {\Big( \frac{k}{k_{BW}} \Big) }^{-2}}, & k \ge k_O.
\end{array} \right.
\end{equation}
%When $k$ is close to $k_{BW}$, $C\Big( \frac{k}{k_{BW}} \Big)$ is close to constant $0.5$ and $E(k) \propto k^{- \frac{5}{3}}$. 
When $k_b<k<k_O$, letting $A=\frac{5\alpha^{\frac{2}{3}}}{12} \frac{v_m}{v_0} \Big( \frac{1-R_f}{R_f}   \Big) $ we obtain:
\begin{eqnarray}
{\Bigg[ 1-AC \Big( \frac{k}{k_{BW}} \Big) \Bigg]}^2 & = & {\sigma_0}^2-2An \sigma_0 \frac{k}{k_{BW}}+ A^2 n^2 { \Big( \frac{k}{k_{BW}} \Big)}^2 + \frac{2}{3}An \sigma_0 {\Big( \frac{k}{k_{BW}} \Big)}^3\nonumber\\ 
&&  +  O \Bigg( {\Big(\frac{k}{k_{BW}} \Big) }^{4} \Bigg)>0,
\end{eqnarray}
where $\sigma_0=A \Big(m+n \arctan \big( \frac{k_O}{k_{BW}}  \big) \Big)-1 $, $m=\frac{\frac{3}{5} {\Big( \frac{k_O}{k_{BW}} \Big) }^{-\frac{5}{3} }} {1+\frac{1}{5} {\Big( \frac{k_O}{k_{BW}} \Big) }^{-2}}$, $n={k_{BW}}^{\frac{2}{3}} {k_O}^{- \frac{2}{3}} $. The spectra slope may vary at different combinations of variables and we will refer to observations.
%
%
%Substituting the above expansion into $E(k)$ and approximating with the first two terms, we further have $E(k)=\alpha \epsilon_0^{\frac{2}{3}} \Big( k^{-\frac{5}{3}} -1.2A k_{BW}^{\frac{5}{3}} k^{-\frac{10}{3}}    \Big),$
%\begin{equation}
%E(k)=\alpha \epsilon_0^{\frac{2}{3}} \Big( k^{-\frac{5}{3}} -1.2A %k_{BW}^{\frac{5}{3}} k^{-\frac{10}{3}}    \Big),
%\end{equation}
%showing that $E(k)$ exhibits a spectrum weaker than the $-5/3$ isotropic turbulence spectrum and corresponding to a spectral ``bump" in the intermediate range. 
When $k>k_O$, the condition \change{$Fr^{\frac{1}{2}} \ll 1.118\frac{v_m}{U} \ll 1$}{$Fr_h^{\frac{1}{2}} \ll 1.118\frac{v_m}{U} \ll 1$} ensures $\frac{k}{k_{BW}} \gg 1$ so $ E(k)=\alpha \epsilon_0^{\frac{2}{3}} k^{-\frac{5}{3}}$ is always \add{a} good approximation and thus the spectrum becomes isotropic. 
%For larger eddies, with $k_b<k<k_{BW}$,
%\begin{eqnarray}
%{\Bigg[ 1-AC \bigg( \frac{k}{k_{BW}} \bigg) \Bigg]}^2 & = & {\sigma}^2 -2.0A\sigma %{\Big(\frac{k}{k_{BW}} \Big) }+A^2{\Big(\frac{k}{k_{BW}} \Big) %}^{2}+\frac{2}{3}a\sigma {\Big(\frac{k}{k_{BW}} \Big) }^{3}  \nonumber\\  
%&& +O \Bigg( {\Big(\frac{k}{k_{BW}} \Big) }^{4} \Bigg),
%\end{eqnarray}  
%where $\sigma=(\frac{\pi}{4}+\frac{1}{2})A-1$. Spectra slope may vary at different combinations of variables and we will refer to observations.
For the largest eddies, $k<k_b$, the condition \change{$Fr^{\frac{1}{2}} \ll 1.118 \frac{v_m}{U} \ll 1$}{ $Fr_h^{\frac{1}{2}} \ll 1.118 \frac{v_m}{U} \ll 1$} ensures that $\frac{k}{k_{BW}} \ll 1$, so $E(k)=\alpha \epsilon_0 {\sigma_0}^2 k^{-\frac{5}{3}}$ is again a good approximation and thus we would expect the spectrum to exhibit a shape similar to the isotropic one but for different physical reasons. 

Therefore, we expect $E(k)$ spectra \add{in horizontal wavenumber} to exhibit a $-5/3$ power-law scaling both below $k_b$ and above $k_O$, and having a scaling that may be different from $-5/3$ in the transition region between $k_b$ and $k_O$. 
%Assuming that the temperature spectrum $E_{\theta}(k)$ is proportional to $E(k)$ \citep{weinstock1985theory}, we have $E_{\theta}(k)=F(N,\epsilon_0)E(k)$,
%\begin{equation}
%E_{\theta}=F(N,\epsilon)E(k),
%\end{equation}
%where $F(N,\epsilon_0)$ is a proportionality faction related to $N$ and $\epsilon_0$ for stratified turbulence and thus we would expect overall similar spectrum behavior for the temperature spectrum. To confirm our theoretical expectation, we now turn to observations and $\mathrm{DNS}$.

\subsubsection{Case $k_a > k_O$}   
If $k_a > k_O$,  similarly to \cite{weinstock1978theory}, we have 
\begin{equation}
C\Big( \frac{k}{k_{BW}} \Big)= \left\{
\begin{array}{ll}
\frac{\frac{3}{5} {\Big( \frac{k_a}{k_{BW}} \Big) }^{-\frac{5}{3} }} {1+\frac{1}{5} {\Big( \frac{k_a}{k_{BW}} \Big) }^{-2}}+{\Big( \frac{k_{BW}}{k_a} \Big) }^{\frac{2}{3} } \Big[ \arctan \Big( \frac{k_a}{k_{BW}} \Big) -\arctan \Big( \frac{k}{k_{BW}}  \Big)  \Big], & k < k_a, \\[4pt]
\frac{\frac{3}{5} {\Big( \frac{k}{k_{BW}} \Big) }^{-\frac{5}{3} }} {1+\frac{1}{5} {\Big( \frac{k}{k_{BW}} \Big) }^{-2}}, & k \ge k_a.
\end{array} \right.
\end{equation}
%If $k_O \le k_a$, scales below Dougherty-Ozmidov scale will be in the isotropic inertial subrange. 
When $k_{b}<k< k_a$,  letting $A=\frac{5\alpha^{\frac{2}{3}}}{12} \frac{v_m}{v_0} \Big( \frac{1-R_f}{R_f}   \Big) $ we obtain:
\begin{eqnarray}
{\Bigg[ 1-AC \Big( \frac{k}{k_{BW}} \Big) \Bigg]}^2 & = & {\sigma_1}^2-2An \sigma_1 \frac{k}{k_{BW}}+ A^2 n^2 { \Big( \frac{k}{k_{BW}} \Big)}^2 + \frac{2}{3}An \sigma_1 {\Big( \frac{k}{k_{BW}} \Big)}^3\nonumber\\ 
&&  +  O \Bigg( {\Big(\frac{k}{k_{BW}} \Big) }^{4} \Bigg)>0,
\end{eqnarray}
where $\sigma_1=A \Big(m+n \arctan \big( \frac{k_a}{k_{BW}}  \big) \Big)-1 $, $m=\frac{\frac{3}{5} {\Big( \frac{k_a}{k_{BW}} \Big) }^{-\frac{5}{3} }} {1+\frac{1}{5} {\Big( \frac{k_a}{k_{BW}} \Big) }^{-2}}$, $n={k_{BW}}^{\frac{2}{3}} {k_a}^{- \frac{2}{3}} $.
%
%
%Substituting the above expansion into $E(k)$ and approximating with the first two terms, we further have $E(k)=\alpha \epsilon_0^{\frac{2}{3}} \Big( k^{-\frac{5}{3}} -1.2A k_{BW}^{\frac{5}{3}} k^{-\frac{10}{3}}    \Big),$
%\begin{equation}
%E(k)=\alpha \epsilon_0^{\frac{2}{3}} \Big( k^{-\frac{5}{3}} -1.2A %k_{BW}^{\frac{5}{3}} k^{-\frac{10}{3}}    \Big),
%\end{equation}
%showing that $E(k)$ exhibits a spectrum weaker than the $-5/3$ isotropic turbulence %spectrum and thus corresponds to a spectral ``bump". When $k>k_O$, the condition $Fr \ll %1.118\frac{v_m}{U} \ll 1$ ensures $\frac{k}{k_{BW}} \gg 1$ so $ E(k)=\alpha %\epsilon_0^{\frac{2}{3}} k^{-\frac{5}{3}}$ is always good approximation and thus the %spectrum becomes isotropic. 
The spectra slope in this case is uncertain and we will illustrate this scenario with observational results. 
When $k>k_a$, the condition \change{$Fr^{\frac{1}{2}} \ll 1.118\frac{v_m}{U} \ll 1$}{$Fr_h^{\frac{1}{2}} \ll 1.118\frac{v_m}{U} \ll 1$} ensures $\frac{k}{k_{BW}} \gg 1$ so $ E(k)=\alpha \epsilon_0^{\frac{2}{3}} k^{-\frac{5}{3}}$ is always \add{a} good approximation and thus the spectrum becomes isotropic. 
For the largest eddies, $k<k_b$, the condition \change{$Fr^{\frac{1}{2}} \ll 1.118 \frac{v_m}{U} \ll 1$}{$Fr_h^{\frac{1}{2}} \ll 1.118\frac{v_m}{U} \ll 1$} ensures that $\frac{k}{k_{BW}} \ll 1$, so $E(k)=\alpha \epsilon_0 {\sigma_1}^2 k^{-\frac{5}{3}}$ is again a good approximation.

%When $ k_{BW} <k < \min (k_O, k_a)$,
%\begin{eqnarray}
%{\Bigg[ 1-AC \Big( \frac{k}{k_{BW}} \Big) \Bigg]}^2 & = & {\sigma_1}^2+2An \sigma_1 { %\Big( \frac{k}{k_{BW}} \Big)}^{-1}+ A^2 n^2 { \Big( \frac{k}{k_{BW}} \Big)}^{-2} - %\frac{2}{3}An \sigma_1 {\Big( \frac{k}{k_{BW}} \Big)}^{-3}\nonumber\\ 
%&&  +  O \Bigg( {\Big(\frac{k}{k_{BW}} \Big) }^{-4} \Bigg)>0.
%\end{eqnarray}

%If $k_O \le k_a$ and when $k_O \le k \le k_a$, the above equation still holds but
% $E(k) \propto k^{-5/3} $ as k is much larger than ${k_{BW}}$. When $k>k_a$, we expect isotropic turbulence due to previous analysis.

%If $k_O > k_a$ and when $k_a \le k \le k_O$, we have
%{\Bigg[ 1-AC \Big( \frac{k}{k_{BW}} \Big) \Bigg]}^2 & = & 1-1.2A  {\Big(\frac{k}{k_{BW}} \Big) }^{-\frac{5}{3}}+0.36A^2 {\Big(\frac{k}{k_{BW}} \Big) }^{-\frac{10}{3}}+\nonumber\\ 
%&& 0.24A  {\Big(\frac{k}{k_{BW}} \Big) }^{-\frac{11}{3}} + O \Bigg( %{\Big(\frac{k}{k_{BW}} \Big) }^{-\frac{16}{3}} \Bigg)>0.
%\end{eqnarray}
%There will thus be a spectra bump between $k_a$ and $k_O$. When $k>k_O$, we again %expect isotropic turbulence. 

To summarize, we expect the following three regions of turbulence spectra in \add{horizontal wavenumber of} the equilibrium range in the stable ABL: the isotropic inertial subrange at $k>\max(k_O,k_a)$; the transition region at $k_b<k<\max (k_O,k_a)$;  the buoyancy subrange with the -5/3 power-law scaling at $k<k_b$. \add{Also note that we do not assume vertical wavenumber spectra are comparable to horizontal wavenumber spectra as only the latter is considered in the ABL.}
%Therefore, we expect $E(k)$ spectra to exhibit a $-5/3$ power-law scaling both below $k_b$ and above $k_O$, and having a scaling that is less steep than $-5/3$ in the transition region between $k_b$ and $k_O$. 
Assuming that the temperature spectrum $E_{\theta}(k)$ \add{in horizontal wavenumber} is proportional to $E(k)$ \citep{weinstock1985theory}, we have $E_{\theta}(k)=F(N,\epsilon_0)E(k)$,
%\begin{equation}
%E_{\theta}=F(N,\epsilon)E(k),
%\end{equation}
where $F(N,\epsilon_0)$ is a proportionality faction related to $N$ and $\epsilon_0$ for stratified turbulence and thus we would expect overall similar spectrum behavior for the temperature spectrum. To confirm our theoretical expectation, we now turn to results from observations and $\mathrm{DNS}$.

%The spectra shape could be applied to TKE spectrum and streamwise velocity spectrum in the horizontal direction of atmospheric boundary layer. However, for the vertical velocity spectrum in the horizontal direction, we need to take into consideration of wall effects as Weinstock's hypothesis does not consider wall effects. As shown in \cite{katul2014two}, vertical velocity spectrum shows a demarcation at $k_a$ where $k_a z=1$.

\section{Experiment and numerical results}\label{sec:Experiment and numerical results}

     \subsection{Observations of the stable atmospheric boundary layer}
     %\subsection{Lake $\mathrm{EC}$ data}
High frequency ($20$ $\mathrm{Hz}$) velocity and temperature were recorded at $4$ different heights ($1.66$ $\mathrm{m}$, $2.31$ $\mathrm{m}$, $2.96$ $\mathrm{m}$ and $3.61$ $\mathrm{m}$ above water level) with eddy-covariance (EC) systems in the stable ABL over Lake Geneva during August-October, $2006$ \citep{bou2008scale}. \add{Wind velocity measurements had errors on the order of $0.021$ $\mathrm{m \: s^{-1}}$ with a maximum of $0.054$ $\mathrm{m \: s^{-1}}$ under zero wind conditions} \citep{vercauteren2008subgrid}. \add{Wind velocity had a standard derivation of $0.001$ $\mathrm{m \: s^{-1}}$ from instrument. Mean of temperature measurements were corrected by the relative mean offsets of instruments. Temperature measurements had a standard deviation of}  \SI{0.002}{\celsius}. \add{The Kolmogorov scales in the $9$ representative periods were $0.0019$ $\mathrm{m}$, $0.0017$ $\mathrm{m}$, $0.0015$ $\mathrm{m}$, $0.0013$ $\mathrm{m}$, $0.0011$ $\mathrm{m}$, $0.0009$ $\mathrm{m}$, $0.0011$ $\mathrm{m}$, $0.0011$ $\mathrm{m}$ and $0.0007$ $\mathrm{m}$ respectively. The smallest scales resolved using Taylor's frozen turbulence hypothesis} \citep{taylor1938spectrum} \add{in the $9$ periods were $0.20$ $\mathrm{m}$, $0.23$ $\mathrm{m}$, $0.25$ $\mathrm{m}$, $0.28$ $\mathrm{m}$, $0.37$ $\mathrm{m}$, $0.50$ $\mathrm{m}$, $0.28$ $\mathrm{m}$, $0.30$ $\mathrm{m}$ and $0.65$ $\mathrm{m}$ respectively. Mean streamwise wind were $1.90$ $\mathrm{m \: s^{-1}}$, $2.27$ $\mathrm{m \: s^{-1}}$, $2.39$ $\mathrm{m \: s^{-1}}$, $2.70$ $\mathrm{m \: s^{-1}}$, $3.62$ $\mathrm{m \: s^{-1}}$, $4.83$ $\mathrm{m \: s^{-1}}$, $2.72$ $\mathrm{m \: s^{-1}}$, $2.92$ $\mathrm{m \: s^{-1}}$ and $6.30$ $\mathrm{m \: s^{-1}}$ respectively. The standard deviation of wind speed were $0.183$ $\mathrm{m \: s^{-1}}$, $0.279$ $\mathrm{m \: s^{-1}}$, $0.369$ $\mathrm{m \: s^{-1}}$, $0.547$ $\mathrm{m \: s^{-1}}$, $0.645$ $\mathrm{m \: s^{-1}}$, $0.648$ $\mathrm{m \: s^{-1}}$, $0.342$ $\mathrm{m \: s^{-1}}$, $0.343$ $\mathrm{m \: s^{-1}}$ and $1.058$ $\mathrm{m \: s^{-1}}$ respectively.} Details about the experiment setup and data can be found in \cite{bou2008scale}, \cite{vercauteren2008subgrid}, \cite{li2011coherent}, and \cite{li2018signatures}.

   High frequency ($10$ $\mathrm{Hz}$) velocity and temperature were recorded at $3.5$ $\mathrm{m}$ above ground with EC in Dome C, Antarctica \citep{vignon2017momentum,vignon2017stable}. The temperature gradient in the stable boundary layer was obtained from balloon sounding measurements \citep{Petenko2018stable}. Four $30$-minute periods around $8$ PM January $9$th, $2015$ were selected. \add{Wind speed measurements had an accuracy of $0.05$ $\mathrm{m \: s^{-1}}$ and temperature had an accuracy of} \SI{0.01}{\celsius} \citep{vignon2017momentum}. \add{The Kolmogorov scales in the 4 representative periods were $0.0022$ $\mathrm{m}$, $0.0015$ $\mathrm{m}$, $0.0023$ $\mathrm{m}$ and $0.0017$ $\mathrm{m}$ respectively. The smallest scales resolved using Taylor's frozen turbulence hypothesis} \citep{taylor1938spectrum} \add{in the 4 periods were $0.45$ $\mathrm{m}$, $0.37$ $\mathrm{m}$, $0.43$ $\mathrm{m}$ and $0.45$ $\mathrm{m}$ respectively. Mean streamwise wind were $2.17$ $\mathrm{m \: s^{-1}}$, $1.81$ $\mathrm{m \: s^{-1}}$, $2.10$ $\mathrm{m \: s^{-1}}$ and $2.16$ $\mathrm{m \: s^{-1}}$ respectively. The standard deviation of wind speed were $0.261$ $\mathrm{m \: s^{-1}}$, $0.182$ $\mathrm{m \: s^{-1}}$, $0.156$ $\mathrm{m \: s^{-1}}$ and $0.215$ $\mathrm{m \: s^{-1}}$ respectively.} Details of the EC setup and the site can be found in \cite{vignon2017momentum}.
   
   Fiber optics were set up at $4$ heights ($1.00$ m, $1.25$ m, $1.50$ m, and $1.75$ m above ground) along a $233$-meter long transect at Oklahoma State University Range Research Station from $2$0 May to $15$ July $2016$. Details of the Distributed Temperature Sensing \add{(DTS)} \citep{selker2006fiber,tyler2009environmental} experiment can be found in \cite{cheng2017failure}. Temperature data were collected every $0.127$ meters along the fiber and every $1.5$ seconds. Mean wind velocity $U$ and scaling temperature $T_*=-\frac{\overline{wT'}}{u_*}$ ($u_*$ is friction velocity) were obtained from a nearby EC tower. Two representative 30-minute periods in stable nocturnal boundary layer were analyzed. \add{Temperature resolution were} \SI{0.16}{\celsius} \add{and} \SI{0.22}{\celsius} \add{before and after the fiber optics transect respectively. The effective spatial resolution was $0.56$ m and temporal resolution was $3$ $\mathrm{s}$. The smallest scales resolved using Taylor's frozen turbulence hypothesis} \citep{taylor1938spectrum} \add{in the $2$ representative periods were both $4.00$ $\mathrm{m}$, while the smallest scales resolved by spatial data were $1.52$ $\mathrm{m}$ in both periods. Mean streamwise wind were both $1.28$ $\mathrm{m \: s^{-1}}$. The standard deviation of temporal temperature in the $2$ periods were} \SI{0.43}{\celsius} \add{and} \SI{0.47}{\celsius} \add{respectively. The standard deviation of spatial temperature of the $2$ periods were} \SI{0.45}{\celsius} \add{and} \SI{0.51}{\celsius}  \add{respectively.}
   
\add{White noise in instruments will typically cause a flat spectrum, so high wavenumber spectrum will be influenced more. In the 2 subplots of the TKE spectra (figure 1) and 1 subplot of temperature spectra (figure 3) in the Lake EC data, a shallower slope than -5/3 can be seen at the highest wavenumbers, which indicates the influence of white noise. In the spectra plots of the Dome C data, the influence of white noise is not obvious. In temporal spectra of DTS data, temperature variation inside the constant temperature cooler box is regarded as white noise, which is then removed from the raw temperature spectra (Schilperoort 2017, personal communication).}

     \subsection{$\mathrm{DNS}$ of stably stratified Ekman layer }
The incompressible Navier-Stokes equations with Boussinesq approximation and the temperature equation are numerically integrated in time. The flow is driven by a steady pressure gradient, assuming a geostrophic balance in regions far above the surface, where the Coriolis force balances the large scale pressure gradient. Numerical details of the code can be found in \cite{shah2014direct}. A neutrally stratified turbulent Ekman layer flow \citep{coleman1992direct} over a smooth surface is first simulated for $ft\approx3$ with $Re_D = U_g D/\nu$, where $f$ the Coriolis parameter, $t$ is time, $U_g$ is the geostrophic wind speed, $D=\sqrt{2\nu/f}$ is the laminar Ekman-layer depth and $\nu$ is the viscosity of air. A cooling surface buoyancy flux, $B_0$, constant in time (i.e. a Neumann boundary condition), is then applied, similarly to \cite{gohari2017direct}.

  A stably stratified Ekman flow is often used to represent an idealized stable planetary boundary layer \citep{ansorge2014global}. The bulk Richardson number $Ri_B = g\delta_*\frac{T_{ref}-T_0}{T_{ref}}$ evolves with time, where  $\delta_*$ is the turbulent Ekman layer length scale given by $\delta_* = u_*/f$, $T_{ref}$ is the reference temperature at far distance and $T_0$ is the surface temperature which changes with time as a result of the imposed cooling buoyancy flux. We can use the Obukhov length scale $L$ \citep{obukhov1946turbulence} to measure the near-surface stability, which is given by $L^+$ when scaled with the inner variables $L^+ =-\frac{u_*^3}{\frac{\kappa g}{\overline{T}} \overline{wT'}  } \frac{u_*}{\nu}$,           
%\begin{equation}
%L_{ob}^+ =(\ - \frac{u_*^3}/{\kappa B_0}\frac{u_*}{\nu}),
%\end{equation}
where $\kappa$ is the von K\'arm\'an constant and $u_* = {[\ (\ \nu\frac{\partial u}{\partial z}\big|_0 )\ ^2 +(\ \nu\frac{\partial v}{\partial z}\big|_0 )\ ^2  ]}^{\frac{1}{4}}$. Using $u_*$ computed from the neutrally stratified case before $B_0$ is applied, we define the initial $L^+(t=0)$ as a measure of the strength of stratification. After imposing $B_0$, the simulation is run for $ft =0.5$ as we would like to mimic the situation of a sudden cooling of the surface in the atmosphere. Table \ref{tab:JFMtable1} shows more details of the setup.

    \subsection{Turbulence spectra \add{in horizontal wavenumber}}
The stability of the atmosphere is related to Obukhov length $L=-\frac{u_*^3}{\frac{\kappa g}{\overline{T}} \overline{wT'}  }$ \citep{obukhov1946turbulence,monin1954basic}, where $u_*$ is friction velocity. Frequency spectra were transformed into \add{horizontal} wavenumber spectra invoking Taylor's frozen turbulence hypothesis \citep{taylor1938spectrum}. Wavelet spectra \citep{torrence1998practical} of temporal series  (the wavelet software was provided by C. Torrence and G. Compo, and is available at URL: http://paos.colorado.edu/research/wavelets/) were calculated.

\begin{figure}
	\centerline{\includegraphics[width=16cm, height=8cm]{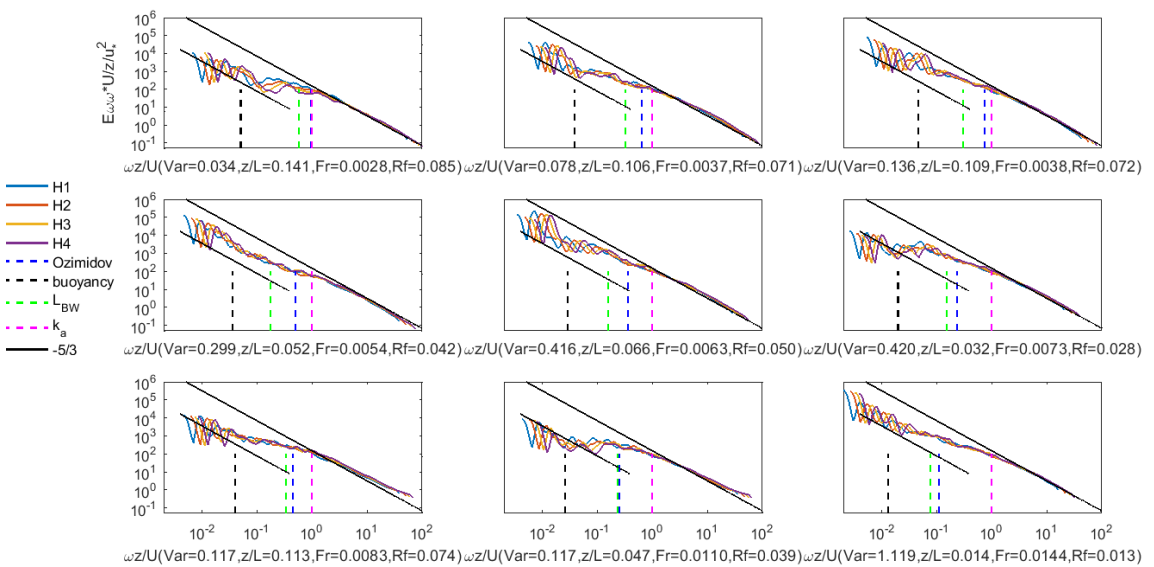}}% Images in 100% size
	\caption{Normalized temporal spectra of $\mathrm{TKE}$ in 9 representative 15-minute periods of Lake Geneva EC data. $E_{\omega\omega}$ is wavelet spectrum in frequency, $U$ is mean streamwise wind velocity, $z$ is measurement height above lake, $u_*$ is friction velocity, $\omega=2\upi f$ is angular frequency, $Var$ is variance of wind velocity, $L$ is Obukhov length, $Fr$ is horizontal Froude number and $R_f$ is flux Richardson number. ``H1”, ``H2", ``H3" and ``H4" denotes observation heights 1.66 m, 2.31 m, 2.96 m and 3.61 m above the lake respectively. ``Ozmidov", ``buoyancy", ``$L_{BW}$" and ``$k_a$" denotes Dougherty-Ozmidov scale, buoyancy scale, $L_{BW}$ and $k_a$ respectively.}
	\label{fig:JFM1}
\end{figure}

\begin{figure}
	\centerline{\includegraphics[width=14cm, height=8cm]{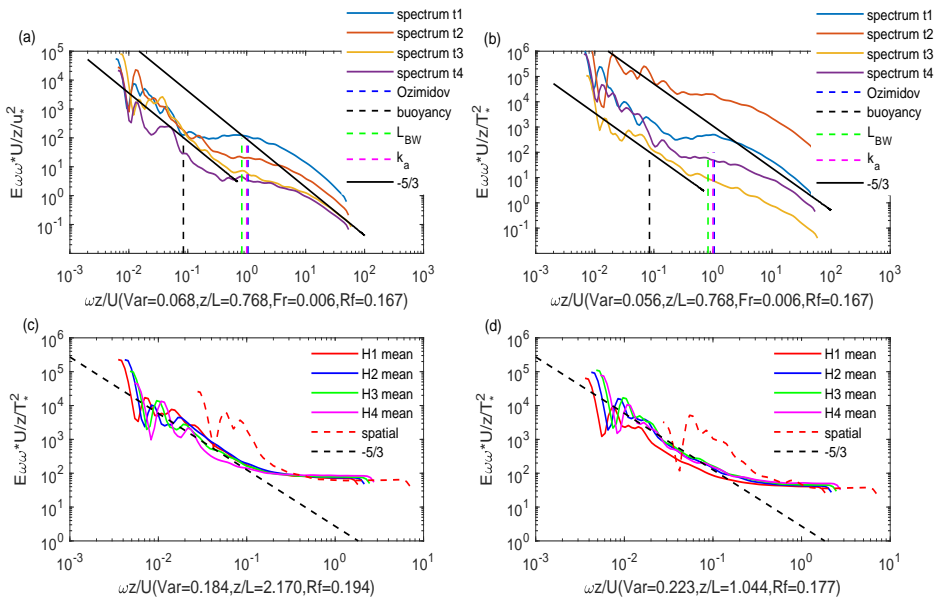}}% Images in 100% size
	\caption{(a) Normalized temporal spectra of $\mathrm{TKE}$ in Dome C data (b) Normalized temporal spectra of temperature in the Dome C data. $T_*$ is scaling temperature and other variables have the same meaning as those in figure 1. ``spectrum t1", ``spectrum t2", ``spectrum t3" and ``spectrum t4" denotes the spectra of 4 different 30-minute stable periods in January 9, 2015 respectively. The 4 length scales correspond to the time period of ``spectrum t1". (c) and (d) \add{Spatial spectra and} spatial mean of temporal spectra of temperature along fiber optics in 2 representative 30-minute periods of $\mathrm{DTS}$ data. Only the buoyancy subrange and transition region are resolved due to limited temporal \add{and spatial} resolution and temporal averaging. Variables have the same meaning as those in figure 1. \add{``spatial" denotes the spatial spectra of temperature at height 1.75 m.}``H1 mean", ``H2 mean", ``H3 mean" and ``H4 mean" denotes the spatial mean of temporal spectra at 4 fiber optics measurement heights respectively.}

	\label{fig:JFM2}
\end{figure}

Similarly to \cite{kaimal1972spectral} both energy spectra \add{in horizontal wavenumber} and frequency are normalized in $\mathrm{log}$-$\mathrm{log}$ plots. In $\mathrm{TKE}$ spectra of Lake EC (figure \ref{fig:JFM1}) and Dome C data (figure \ref{fig:JFM2}a), there are three regions corresponding to the buoyancy subrange, a transition region and the isotropic inertial subrange. At $k>\max(k_O,k_a)$, the spectra match the -5/3 power-law scaling of isotropic turbulence \citep{kolmogorov1941,dougherty1961anisotropy,ozmidov1965turbulent}. The observation also shows that Dougherty-Ozmidov scale is a better demarcation of the isotropic inertial subrange compared to $L_{BW}$ defined in \cite{weinstock1978theory}. In the transition region at $k_b<k<\max(k_O,k_a)$, the slope of spectra is not universal but less steep than $-5/3$. \cite{katul2012existence} has reported a $-1$ slope at low wavenumbers in the $\mathrm{TKE}$ spectra of the  stable atmospheric boundary layer. It appears to be a specific case of the transition region described here, with a slope not as steep as $-5/3$. Above the buoyancy scale, observations show another $-5/3$ slope, %which is consistent with Large-Eddy Simulation (LES) studies \citep{khani2014buoyancy,waite2011stratified}, although the latter used anisotropic turbulence hypothesis
%\citep{lindborg2006energy} to interpret the spectrum above buoyancy scale.  
which agrees with spectra shown in figure 7 of \cite{muschinski2004small} and is consistent with the theory presented in section 2.

%The maximum of the buoyancy scale in these plots is around $6$ km and the minimum around $300$ m. The lowest wavenumber parts in the spectra are larger than $2$ km, which is the upper limit of the universal equilibrium range in \cite{weinstock1978theory}. So eddies corresponding to lowest wavenumber parts in the spectra may be regarded as anisotropic turbulence \citep{lindborg2006energy,waite2011stratified}.

% insert table 1 here
\begin{table}
	\begin{center}
		\def~{\hphantom{0}}
		\begin{tabular}{lcccc}
			$Re_D$  & $L^+(t=0)$   & $L_x,L_y,L_z$ & $\Delta x(y)^+,\Delta z^+$   & $N_x\times N_y\times N_z$  \\[3pt]
			1 000   & 1 600         & $90$D, $90$D, $30$D   & $3.90$, $0.353$       & 1024 $\times$ 1024 $\times$ 3840  \\
		\end{tabular}
		\caption{Details of $\mathrm{DNS}$ setup.  $L_x$, $L_y$ and $L_z$ are the dimensions of computational domain in $x$ (streamwise), $y$ (spanwise) and $z$ (vertical) directions respectively; $\Delta x(y)^+,\Delta z^+$are the resolutions in $x(y)$ and $z$ directions; $N(x, y, z)$ denotes the number of points for computation.}
		\label{tab:JFMtable1}
	\end{center}
\end{table}

% insert Figure 1 here  %scale=0.5 width=16cm, height=8cm

% insert Figure 2 here   width=14cm, height=6cm  scale=0.40

\begin{figure}
	\centerline{\includegraphics[width=16cm, height=8cm]{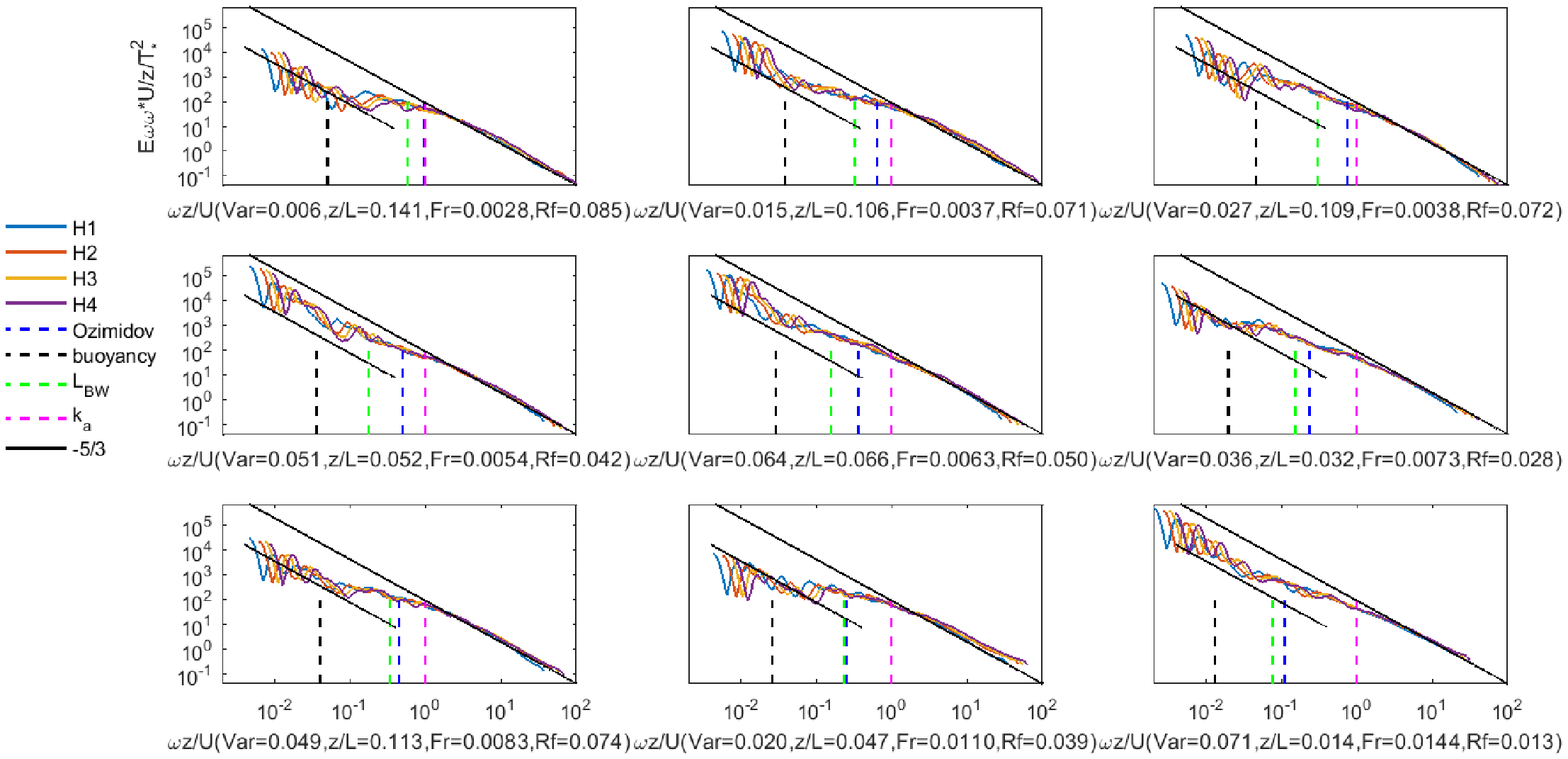}}% Images in 100% size
	\caption{Temporal spectra of temperature in 9 representative 15-minute periods of Lake EC data. $T_*$ is scaling temperature and other variables have the same meaning as those in figure 1.}
	\label{fig:JFM3}
\end{figure}

In the temperature spectra of the Dome C (figure \ref{fig:JFM2}b) and Lake data (figure \ref{fig:JFM3}), three regimes similar to the $\mathrm{TKE}$ spectrum can be observed. The observed $-1$ slope \citep{katul2016deviations,li2015turbulent} at low wavenumber in the temperature spectra of stable atmospheric boundary layer is actually a specific scaling in the transition region.
%
%
%
%
% insert Figure 3 here scale=0.50
In the temperature spectra of $\mathrm{DTS}$ data (figure \ref{fig:JFM2}c and \ref{fig:JFM2}d), only the buoyancy subrange and transition region are observed due to limited temporal \add{and spatial} resolution and temporal averaging \citep{cheng2017failure}. \add{The spatial spectra were used to check if temporal spectra that invoke Taylor's frozen turbulence hypothesis are a good approximation, but those spatial spectra are more limited in terms of the number of decades they cover ($\sim 3$). More detailed discussion on Taylor's hypothesis can be found in} \cite{cheng2017failure}. \add{Both temporal spectra and spatial spectra exhibit an approximately -5/3 slope at low wavenumber and a shallower slope at high wavenumber.} The minimum length resolved in the \change{$\mathrm{DTS}$}{temporal and spatial} spectra \change{is}{were}\remove{about} \change{$3$}{$4.00$} \add{and $1.52$} meters respectively, which are close to the Dougherty-Ozmidov scale typically observed in the atmospheric boundary layer \citep{jimenez2005large,li2016connections} so that we cannot observe the smaller scale isotopic behavior. About one decade to the left of the Dougherty-Ozmidov scale, a $-5/3$ scaling can be observed, corresponding to the buoyancy subrange previously defined. The transition region in the $\mathrm{DTS}$ data resembles a flat white noise spectrum compared to other observations. One possible explanation is that spatial averaging is applied to the temporal spectrum of a single spatial point. 
%
% insert Figure 4 here
%\begin{figure}
%	\centerline{\includegraphics[scale=0.4]{JFMf4new}}% Images in 100% size
%	\caption{Spatial mean of temporal spectra of T along fiber optics in 4 representative 30-minute periods of $\mathrm{DTS}$ data. Only the buoyancy subrange and transition region are resolved due to limited temporal resolution. Variables have the same meaning as those figure 1. ``H1 mean", ``H2 mean", ``H3 mean" and ``H4 mean" denotes the spatial mean of temporal spectra at 4 fiber optics measurement heights respectively.}
%	\label{fig:JFM4}
%\end{figure}
%
%

\begin{figure}
	\centerline{\includegraphics[width=14cm, height=5cm]{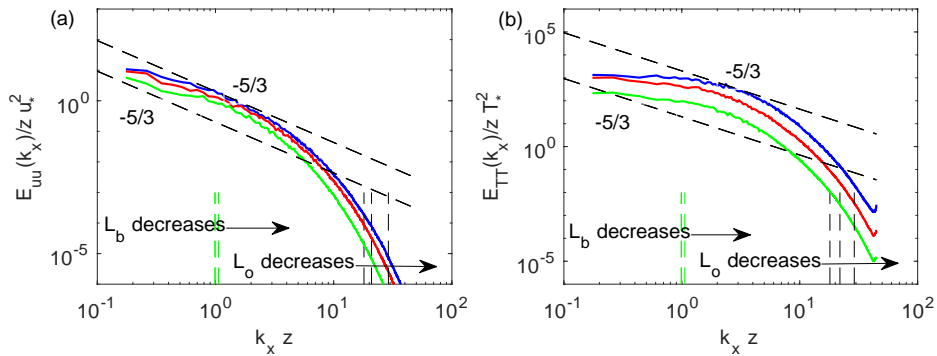}} %{spectra_u_T_dbres}} %{spectra_uandT_new}}% Images in 100% size
	\caption{One-dimensional streamwise spectra scaled similarly to the observational data: (a) One-dimensional streamwise spectra of \change{u}{$u$}. (b) One-dimensional streamwise spectra of \change{T}{$T$}. Blue, red and green lines indicate the spectra at $ft$=0.02, 0.14 and 0.20 with respective increasing bulk $Ri_B$  from 0.21, 0.65 to 0.98 taking at approximately $z^+=65$ to fall in the inertial layer. The blacked dotted lines indicate the classic $-5/3$ slope; the vertical green (black) dotted lines denote the scaled wavenumber corresponding to the decreasing buoyancy length scale $L_b$ (Dougherty-Ozmidov length scale $L_O$). Variables have the same meaning as those in figure 1.}
	\label{fig:JFM5}
\end{figure}

The Kolmogorov scale \remove{shown}\add{(see introduction in section $3.1$)} in Lake EC data is about 4 decades away from the Dougherty-Ozmidov scale, suggesting that turbulence is still well defined rather than being intermittent as shown in $\mathrm{DNS}$ results \citep{flores2011analysis,ansorge2016analyses}. The spectral bump (figure \ref{fig:JFM5}) as $Ri_B$ changes from $0.21$, $0.65$ to $0.98$ in the $\mathrm{DNS}$ is not as clear at those shown in the atmospheric data. Indeed, in the $\mathrm{DNS}$ the Dougherty-Ozmidov scale is nearly similar to Kolmogorov scale \citep{waite2014direct}. It may be argued that around the $k_x z$  indicated by the scale of $1/L_b$, the spectral bump appears consistent with the scaling analysis above. However, $\mathrm{DNS}$ can only achieve a narrow range of inertial subrange compared to the observational data in the atmospheric boundary layer, which is at a much higher Reynolds number and thus $\mathrm{DNS}$ do not appear to have sufficient scale separation \citep{kunkel2006study} to correctly represent  the three regimes highlighted above \add{due to the limitation of current computational capacity}. As stratification increases, the Dougherty-Ozmidov length scale further decreases (and the wavenumber $k_O$  increases), the highest wavenumbers become closer to the $-5/3$ scaling and are less impacted by dissipative effects. However, a larger separation of scales between the inertial subrange and the dissipation range is required and not met with currently achievable $\mathrm{DNS}$ Reynolds numbers. 
Although $\mathrm{DNS}$ has been useful in reproducing quantitatively and qualitatively some surface layer similarity relations according to MOST \citep{chung2012direct}, we have to question the applicability of $\mathrm{DNS}$ for the study of very stable boundary layers \add{at very high Reynolds numbers} as they cannot correctly represent the observed spectral regimes seen in actual atmospheric boundary layers because of the lack of scale separation \add{due to the limitation of current computational capacity}. $\mathrm{LES}$ will obviously face similar issues in addition to the subgrid scale modeling that will have to be applied at scales well below the Dougherty-Ozmidov length scale to be applicable. In $\mathrm{LES}$, spectra below the Dougherty-Ozmidov scale are thus not present \citep{beare2006intercomparison,waite2011stratified,khani2014buoyancy}.
%
% insert Figure 5 here  width=12cm, height=5cm scale=0.39
% insert Figure 6 here scale=0.39
\begin{figure}
	\centerline{\includegraphics[width=14cm, height=7cm]{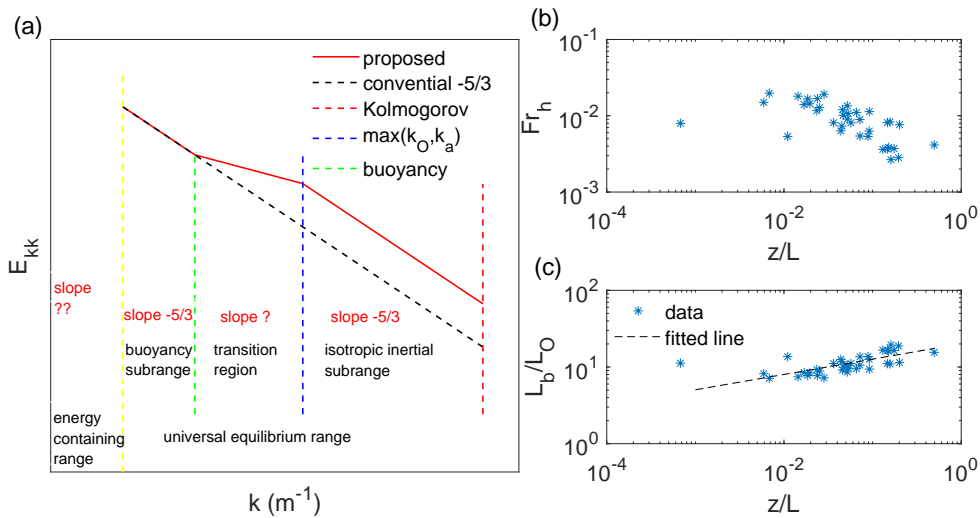}}% Images in 100% size
	\caption{(a) Schematic of $\mathrm{TKE}$ and temperature spectra in horizontal wavenumber in the stable atmospheric boundary layer. (b) The scatter plot between horizontal Froude number \change{$Fr$}{$Fr_h$} and $z/L$ of Lake EC data. (c) The scatter plot between ratio of buoyancy scale $L_b$ to Dougherty-Ozmidov scale $L_O$ and $z/L$ of Lake EC data.}
	\label{fig:JFM6}
\end{figure}

A schematic of the $\mathrm{TKE}$ and temperature spectra \add{in horizontal wavenumber} in the stable atmospheric boundary layer is shown (figure \ref{fig:JFM6}a). In the universal equilibrium range, the -5/3 power-law scaling at scales larger than buoyancy scale is due to stratification effects \citep{weinstock1978theory}, while the -5/3 scaling at $k>\max(k_O,k_a)$ is due to isotropic turbulence
\citep{dougherty1961anisotropy,kolmogorov1941,ozmidov1965turbulent}. In the intermediate transition region, there is not a universal power-law scaling for spectra although the slope is less steep than $-5/3$ and even sometimes close to white noise. Our focus is the equilibrium range in the atmospheric boundary layer so spectra slope at the energy containing range is not shown. %At very large scales above the equilibrium range, we found a scaling close to $-5/3$ in anisotropic turbulence in the Lake EC and Dome C data, which agrees with the anisotropic hypothesis \citep{lindborg2006energy}. 
%For temperature spectra, a $-3$ scaling has been proposed for anisotropic turbulence \citep{phillips1965bolgiano,weinstock1985theory,holloway1986considerations} even though it was not observed in our case. 
%Therefore, more research are needed to better determine the spectra in energy containing range.
%

The Lake EC data also show (figure \ref{fig:JFM6}b) that \change{$Fr$}{$Fr_h$} and $z/L$ are consistent in describing stability of atmospheric boundary layer, i.e., the higher $z/L$, the lower \change{$Fr$}{$Fr_h$} and the more stable the atmosphere is. When $k_O>k_a$, the width of the transition region denoted by the ratio of buoyancy scale \change{and}{to} Dougherty-Ozmidov scale increases with $z/L$ (figure \ref{fig:JFM6}c). Therefore, the transition region will play a more important role in more stable conditions. Due to the increased impact of the transition region, the typically assumed $-5/3$ spectra scaling \citep{kaimal1972spectral} has to be revised to better represent $\mathrm{TKE}$, temperature variance or other fluxes while applying MOST. In the limit of extremely stable boundary layer, the transition region would occupy a very large fraction of the inertial subrange as the Dougherty-Ozmidov scale approaches the Kolmogorov scale, which may be interpreted as a collapse of turbulence. However, such collapse of turbulence was not observed in our data in the atmospheric boundary layer.

    \subsection{Discussion}
In \cite{kaimal1973turbulenece}, a -5/3 power-law scaling was shown in the spectra of $u$, $v$ and $w$ at the high-frequency \add{horizontal} wavenumbers and was approximated by an empirical formula. The shallower slope of spectra in the ``transition region'' between the buoyancy scale and Dougherty-Ozmidov scale was not reported. However, figure 1 in \cite{caughey1977boundary} showed a shallower slope in $u$ and $T$ spectra at lower frequency compared to the isotropic -5/3 scaling frequency range at the height of 8 meters. Such a shallower slope of spectra was not as obvious at heights of 46 meters and 91 meters above the surface, at which the atmosphere was not as stable as at 8 meters. So figure 1 in \cite{caughey1977boundary} actually showed the existence of a transition region to the left hand side of the isotropic -5/3 scaling in log-log plots. Figure 2 in \cite{caughey1977boundary} showed the slope of the transition region was shallower than -5/3. The Dougherty-Ozmidov scale and buoyancy scale were not shown in figures of \cite{caughey1977boundary} so the importance of the two scales under stratification might have been overlooked at that time. \cite{grachev2015similarity} proposed a similarity theory based on the Dougherty-Ozmidov scale under stable conditions, which provided the evidence that Dougherty-Ozmidov characterizes the small scale turbulence well. \cite{li2016connections} showed Dougherty-Ozmidov scale was the limitation of momentum transporting eddies as stability increased. These results were consistent with our finding that Dougherty-Ozmidov scale decreases and is the limit of isotropic turbulence as stratification increases. \cite{smyth2000length} suggested that the buoyancy Reynolds number is limited in \add{current} $\mathrm{DNS}$ studies \add{due to the computational limitation} and that turbulence decays when bulk Richardson number exceeds 1/4, which agrees with the condition $R_f<0.25$ applied here for continuous Richardson-Kolmogorov cascade. \cite{riley2008stratified} showed that some geophysical turbulence spectra under stratification at scales larger than the Dougherty-Ozmidov scale should not be explained by Kolmogorov's isotropic turbulence hypothesis. We further show that stratification influences the turbulence spectra at scales larger than the Dougherty-Ozmidov scale by observation and derivation. The “transition region” that is highlighted in our work was not emphasized in previous research such as in \cite{riley2008stratified} or \cite{lindborg2006energy}. 

The original theory of \cite{weinstock1978theory} applied the ``locally inertial" relation $E(k)=\alpha \epsilon(k)^{2/3} k^{-5/3}$ for ``approximately isotropic" turbulence in the equilibrium range (scales smaller than energy containing range) on condition that $\big| \frac{k}{\epsilon} \frac{\partial \epsilon}{\partial k} \big|\ll 1$. The condition $\big| \frac{k}{\epsilon} \frac{\partial \epsilon}{\partial k} \big|\ll 1$ is shown to be satisfied in most parts (see pages 644-645 of \cite{weinstock1978theory}), so the ``locally inertial" assumption is approximately valid in that region \add{when stratification is weak}. The deviation of the TKE spectra from isotropy is not large, which was supported by observation \citep{reiter1965atmospheric} that vertical velocity spectrum is comparable to horizontal spectrum. So the theory is self-consistent in assuming ``local inertial" relation in the equilibrium range when \change{there are no walls such as in the stratosphere}{vertical wavenumber spectra are similar to horizontal wavenumber spectra and wall effects can be neglected}. \change{One may argue that Weinstock's theory is inherently isotropic and may not apply to stratified turbulence which is ``anisotropic"}{Indeed, Weinstock's theory is inherently isotropic and does not apply to strongly stratified turbulence which is ``anisotropic"} \add{as horizontal length scale will be much larger than vertical length scale under strong stratification} \citep{billant2001self}. \change{For example}{For strongly stratified turbulence}, people may assume that Lindborg's stratified turbulence hypothesis \citep{lindborg2006energy} rather than Weinstock's theory describes turbulence above Dougherty-Ozmidov scale. However, in the case of homogeneous stratified turbulence in DNS studies, the suggested vertical wavenumber $k_v^{-3}$ scaling by \cite{lindborg2006energy} is not \add{always} reproduced. \cite{waite2004stratified} showed $k_v^{-5/3}$ spectrum in the vertical direction under weak stratification. \cite{waite2006stratified} reported shallower spectrum than  $k_v^{-3}$. \cite{bartello2013sensitivity} suggested no evidence of vertical spectrum $k_v^{-3}$ expected between the buoyancy scale and Dougherty-Ozmidov scale, which is due to limited stratification. \cite{maffioli2016dynamics} also showed the vertical spectra are closer to $k_v^{-5/3}$ rather than $k_v^{-3}$ owing to insufficiently low Froude number. \cite{almalkie2012kinetic} suggested $k_v^{-3}$ is not apparent in most cases of stratified turbulence simulations. \cite{maffioli2017vertical} used a scale decomposition and shows the $k_v^{-3}$ scaling in large scales\remove{and $k_v^{-5/3}$ in small scales}. An approximate balance \citep{maffioli2017vertical} between buoyance and inertia has been proposed for scales between Dougherty-Ozmidov scale and buoyancy scale. These studies suggest Lindborg's $k_v^{-3}$  hypothesis cannot entirely describe vertical wavenumber spectrum around the buoyancy scale and that vertical wavenumber spectra is not that different from horizontal wavenumber spectrum \add{in weakly stratified conditions where horizontal length scale is close to vertical length scale}. Therefore, the “approximately isotropic” hypothesis may still partly apply under certain conditions in \add{weakly} stratified turbulence.

%In the equilibrium range, Weinstock's theory applies to total k spectrum while assuming homogeneous turbulence in the flow field. In the case of atmospheric boundary layer, k3 direction is not homogeneous but k1 and k2 directions could be considered homogeneous. Generally k3 spectrum is not calculated and the information in k3 spectrum is neglected in the atmospheric boundary layer. Therefore, TKE spectrum in k1 or k2 direction in the equilibrium range could be described by Weinstock's theory in the atmospheric boundary layer.

\add{Different from Weinstock (1978), we do not assume that vertical wavenumber spectra are comparable to horizontal wavenumber spectra, i.e., isotropic hypothesis is not applied here. For horizontal wavenumbers, we do not assume a similar spectra shape between $w$ and $u$ in the ABL.} Indeed, in the atmospheric boundary layer, turbulence in the vertical direction (perpendicular to land surface) is not homogeneous \citep{fiedler1970atmospheric} but turbulence in the horizontal directions can be considered homogeneous. Generally vertical wavenumber spectrum is not calculated or considered in the atmospheric boundary layer. For turbulence spectrum at horizontal wavenumber, we relax the isotropic hypothesis and assume $w$ spectrum is different from $u$ spectrum at low wavenumbers.
One may still argue that the $-5/3$ scaling around the buoyancy scale in our observation should be explained by Lindborg's anisotropic $k_h^{-5/3}$ cascade. Compared to free atmosphere motion, atmospheric boundary layer turbulence is microscale motion and is essentially three-dimensional \citep{fiedler1970atmospheric,monin2013statistical} as boundary layer scale is much smaller than the density scale height (about 10 km). \cite{waite2011stratified} suggested that both buoyancy scale and Dougherty-Ozmidov scale are much smaller than energy containing horizontal scale in stratified turbulence. That is to say, scales around the buoyancy scale and Dougherty-Ozmidov scale are in the equilibrium range in the stable atmospheric boundary layer, which supports our analysis for horizontal wavenumber spectrum. \cite{waite2011stratified} also pointed out that the $k_h^{-5/3}$ cascade is driven by anisotropic eddies with horizontal scales much larger than buoyancy scale, i.e., Lindborg's stratified turbulence cascade mainly describes turbulence in the energy containing range. Therefore, horizontal turbulence in the energy containing range is described by Lindborg's $k_h^{-5/3}$ cascade,  while our analysis is below the energy containing range.
Besides, a lot of numerical studies \citep{augier2015stratified,brethouwer2007scaling,maffioli2016dynamics,waite2011stratified} showed a “spectral bump” around the buoyancy scale, which is neither predicted nor explained by Lindborg's $k_h^{-5/3}$ spectra of anisotropic turbulence hypothesis. These results also suggest that buoyancy scale is not in the energy containing range in the horizontal direction and support our equilibrium range analysis.

\add{Here some hypotheses on the spectra shape are discussed.} \cite{townsend1958turbulent} \add{assumed that fluid motion may consist of gravity waves when buoyancy forces are dominant.}  \cite{bolgiano1959turbulent} \add{noted that TKE is converted to potential energy in the case of buoyancy stratification.} \cite{lumley1964spectrum} \add{suggested that wave-like behavior exists at low wavenumbers.} \cite{weinstock1978theory} \add{showed that gravity waves are weakly damped at low wavenumbers but are heavily damped at high wavenumbers in the isotropic inertial subrange.}  \cite{zilitinkevich2007energy} \add{showed energy transfer between TKE and turbulent potential energy from budget equations for both of them.	Similarly to the derivation in} \cite{weinstock1978theory}\add{, the -5/3 spectra slope at scales larger than the buoyancy scale is caused by constant dissipation rate, which is smaller than $\epsilon_0$ in isotropic turbulence below the Dougherty-Ozmidov scale. However, the variation of dissipation rate in wavenumber space leads to different spectra slopes in the transition region. Therefore, the different spectra characteristics between buoyancy subrange and transition region correspond to different interaction strengths between gravity waves and turbulence. The spectra shallower than -5/3 in the transition region indicate extra energy between buoyancy scale and Dougherty-Ozmidov scale. The extra energy might possibly come from potential energy that is converted to kinetic energy through counter-gradient heat flux} \citep{holt1992numerical,komori1996effects,schumann1987countergradient,zilitinkevich2007energy} \add{in stratified turbulence. Another possible explanation of the shallower spectra might be Kelvin-Helmholtz instabilities} \citep{brethouwer2007scaling,waite2011stratified}. \add{In fact,} \cite{brethouwer2007scaling} \add{found counter-gradient fluxes appearing nearly simultaneously with Kelvin-Helmholtz-type instabilities. Therefore, further research are needed to explain the relation between shallower spectra, counter-gradient fluxes and Kelvin-Helmholtz instabilities. However, it is unlikely that an inverse energy cascade causes shallower spectra in transition region because direct transfer of energy was shown} \citep{augier2015stratified,waite2011stratified} \add{from energy-containing scale to buoyancy scale}.
\section{Conclusion}
In the stable atmospheric boundary layer, we showed for the first time that buoyancy subrange, transition region and isotropic inertial subrange are separated by $k_b$ and $max(k_O,k_a)$ in $\mathrm{TKE}$ and temperature spectra of \add{horizontal wavenumber in} the equilibrium range. The transition region between buoyancy scale and Dougherty-Ozmidov scale will be observed when \change{$Fr^{1/2} \ll 1.118 \frac{v_m}{U} \ll 1$}{$Fr_h^{1/2} \ll 1.118 \frac{v_m}{U} \ll 1$} and  $R_f < 0.25$ (for the existence of continuous turbulence with Richardson-Kolmogorov cascade) are satisfied. To represent the full spectra \add{in horiozntal wavenumber} in the \add{stably stratified} atmospheric boundary layer, $\mathrm{LES}$ needs to resolve scales as small as the Dougherty-Ozmidov scale and use suitable subgrid scale model. To correctly represent those regimes, $\mathrm{DNS}$ would have to be run at much higher Reynolds number to obtain larger scale separation between the Dougherty-Ozmidov and Kolmogorov scale, as well as in the energy containing range. The impact of the variation of the spectra in the transition region should generate departure from MOST in very stable atmospheric boundary layer as this region expands when stratification increases.
\\
\begin{flushleft}
\textbf{Acknowledgements}
\end{flushleft}

PG would like to acknowledge funding from the National Science Foundation (NSF CAREER, EAR-1552304), and from the Department of Energy (DOE Early Career, DE-SC00142013). The lake data were collected by the Environmental Fluid Mechanics and Hydrology Laboratory of Professor M. Parlange at L'\`Ecole Polytechnique F\'ed\'erale de Lausanne. We would like to thank Prof. M. Parlange and Prof. Elie Bou-Zeid for sharing Lake EC data and thank Prof. Jeffrey Basara for sharing MOISST EC data. Dome C data were acquired in the frame of the projects “Mass lost in wind flux” (MALOX) and “Concordia multi-process atmospheric studies” (COMPASS) sponsored by PNRA. A special thanks to P. Grigioni and all the staff of “Antarctic Meteorological Observatory” of Concordia for providing the radio sounding used in this study. And a special thank to Dr. Igor Petenko of CNR ISAC for running the field experiment at Concordia station. We would also like to thank Prof. John Selker and Center for Transformative Environmental Monitoring Programs (CTEMPs) for help in the DTS experiment. We would like to acknowledge the National Center of Atmospheric Research computing facilities Yellowstone and Cheyenne where the DNS study was performed.

\appendix

\bibliographystyle{jfm}
\bibliography{jfm-instructions}

\begin{thebibliography}{89}
\expandafter\ifx\csname natexlab\endcsname\relax\def\natexlab#1{#1}\fi
\def\au#1{#1} \def\ed#1{#1} \def\yr#1{#1}\def\at#1{#1}\def\jt#1{\textit{#1}}
  \def\bt#1{#1}\def\bvol#1{\textbf{#1}} \def\vol#1{#1} \def\pg#1{#1}
  \def\publ#1{#1}\def\arxiv#1{#1}\def\org#1{#1}\def\st#1{\textit{#1}}

\bibitem[Almalkie \& de~Bruyn~Kops(2012)]{almalkie2012kinetic}
{\sc \au{Almalkie, Saba} \& \au{de~Bruyn~Kops, Stephen~M}} \yr{2012}
  \at{Kinetic energy dynamics in forced, homogeneous, and axisymmetric stably
  stratified turbulence}.  \jt{Journal of Turbulence} ~(13),  \pg{N29}.

\bibitem[Ansorge \& Mellado(2014)]{ansorge2014global}
{\sc \au{Ansorge, C.} \& \au{Mellado, J.~P.}} \yr{2014}  \at{Global
  intermittency and collapsing turbulence in the stratified planetary boundary
  layer}.  \jt{Boundary-layer meteorology}  \bvol{153}~(1),  \pg{89--116}.

\bibitem[Ansorge \& Mellado(2016)]{ansorge2016analyses}
{\sc \au{Ansorge, Cedrick} \& \au{Mellado, Juan~Pedro}} \yr{2016}  \at{Analyses
  of external and global intermittency in the logarithmic layer of ekman flow}.
   \jt{Journal of Fluid Mechanics}  \bvol{805},  \pg{611--635}.

\bibitem[Augier {\em et~al.\/}(2015)Augier, Billant \&
  Chomaz]{augier2015stratified}
{\sc \au{Augier, Pierre}, \au{Billant, Paul} \& \au{Chomaz, Jean-Marc}}
  \yr{2015}  \at{Stratified turbulence forced with columnar dipoles: numerical
  study}.  \jt{Journal of Fluid Mechanics}  \bvol{769},  \pg{403--443}.

\bibitem[Baker \& Gibson(1987)]{baker1987sampling}
{\sc \au{Baker, Mark~A} \& \au{Gibson, Carl~H}} \yr{1987}  \at{Sampling
  turbulence in the stratified ocean: Statistical consequences of strong
  intermittency}.  \jt{Journal of Physical Oceanography}  \bvol{17}~(10),
  \pg{1817--1836}.

\bibitem[Bartello \& Tobias(2013)]{bartello2013sensitivity}
{\sc \au{Bartello, P} \& \au{Tobias, SM}} \yr{2013}  \at{Sensitivity of
  stratified turbulence to the buoyancy reynolds number}.  \jt{Journal of Fluid
  Mechanics}  \bvol{725},  \pg{1--22}.

\bibitem[Beare {\em et~al.\/}(2006)Beare, Macvean, Holtslag, Cuxart, Esau,
  Golaz, Jimenez, Khairoutdinov, Kosovic, Lewellen {\em
  et~al.\/}]{beare2006intercomparison}
{\sc \au{Beare, Robert~J}, \au{Macvean, Malcolm~K}, \au{Holtslag, Albert~AM},
  \au{Cuxart, Joan}, \au{Esau, Igor}, \au{Golaz, Jean-Christophe}, \au{Jimenez,
  Maria~A}, \au{Khairoutdinov, Marat}, \au{Kosovic, Branko}, \au{Lewellen,
  David} \& \au{others}} \yr{2006}  \at{An intercomparison of large-eddy
  simulations of the stable boundary layer}.  \jt{Boundary-Layer Meteorology}
  \bvol{118}~(2),  \pg{247--272}.

\bibitem[Billant \& Chomaz(2001)]{billant2001self}
{\sc \au{Billant, Paul} \& \au{Chomaz, Jean-Marc}} \yr{2001}
  \at{Self-similarity of strongly stratified inviscid flows}.  \jt{Physics of
  Fluids}  \bvol{13}~(6),  \pg{1645--1651}.

\bibitem[Bolgiano(1959)]{bolgiano1959turbulent}
{\sc \au{Bolgiano, R}} \yr{1959}  \at{Turbulent spectra in a stably stratified
  atmosphere}.  \jt{Journal of Geophysical Research}  \bvol{64}~(12),
  \pg{2226--2229}.

\bibitem[Bou-Zeid {\em et~al.\/}(2008)Bou-Zeid, Vercauteren, Parlange \&
  Meneveau]{bou2008scale}
{\sc \au{Bou-Zeid, E.}, \au{Vercauteren, N.}, \au{Parlange, M.~B.} \&
  \au{Meneveau, C.}} \yr{2008}  \at{Scale dependence of subgrid-scale model
  coefficients: an a priori study}.  \jt{Physics of Fluids}  \bvol{20}~(11),
  \pg{115106}.

\bibitem[Brethouwer {\em et~al.\/}(2007)Brethouwer, Billant, Lindborg \&
  Chomaz]{brethouwer2007scaling}
{\sc \au{Brethouwer, G.}, \au{Billant, P.}, \au{Lindborg, E.} \& \au{Chomaz,
  J-M}} \yr{2007}  \at{Scaling analysis and simulation of strongly stratified
  turbulent flows}.  \jt{Journal of Fluid Mechanics}  \bvol{585},
  \pg{343--368}.

\bibitem[Caughey(1977)]{caughey1977boundary}
{\sc \au{Caughey, SJ}} \yr{1977}  \at{Boundary-layer turbulence spectra in
  stable conditions}.  \jt{Boundary-Layer Meteorology}  \bvol{11}~(1),
  \pg{3--14}.

\bibitem[Charney(1971)]{charney1971geostrophic}
{\sc \au{Charney, Jule~G}} \yr{1971}  \at{Geostrophic turbulence}.  \jt{Journal
  of the Atmospheric Sciences}  \bvol{28}~(6),  \pg{1087--1095}.

\bibitem[Cheng {\em et~al.\/}(2005)Cheng, Parlange \&
  Brutsaert]{cheng2005pathology}
{\sc \au{Cheng, Yinguo}, \au{Parlange, Marc~B} \& \au{Brutsaert, Wilfried}}
  \yr{2005}  \at{Pathology of monin-obukhov similarity in the stable boundary
  layer}.  \jt{Journal of Geophysical Research: Atmospheres}  \bvol{110}~(D6).

\bibitem[Cheng {\em et~al.\/}(2017)Cheng, Sayde, Li, Basara, Selker, Tanner \&
  Gentine]{cheng2017failure}
{\sc \au{Cheng, Y.}, \au{Sayde, C.}, \au{Li, Q.}, \au{Basara, J.}, \au{Selker,
  J.}, \au{Tanner, E.} \& \au{Gentine, P.}} \yr{2017}  \at{Failure of taylor's
  hypothesis in the atmospheric surface layer and its correction for
  eddy-covariance measurements}.  \jt{Geophysical Research Letters}
  \bvol{44}~(9),  \pg{4287--4295}.

\bibitem[Chung \& Matheou(2012)]{chung2012direct}
{\sc \au{Chung, Daniel} \& \au{Matheou, Georgios}} \yr{2012}  \at{Direct
  numerical simulation of stationary homogeneous stratified sheared
  turbulence}.  \jt{Journal of Fluid Mechanics}  \bvol{696},  \pg{434--467}.

\bibitem[Coleman {\em et~al.\/}(1992)Coleman, Ferziger \&
  Spalart]{coleman1992direct}
{\sc \au{Coleman, GN}, \au{Ferziger, JH} \& \au{Spalart, PR}} \yr{1992}
  \at{Direct simulation of the stably stratified turbulent ekman layer}.
  \jt{Journal of Fluid Mechanics}  \bvol{244},  \pg{677--712}.

\bibitem[Derbyshire(1995)]{derbyshire1995stable}
{\sc \au{Derbyshire, SH}} \yr{1995}  \at{Stable boundary layers: Observations,
  models and variability part i: Modelling and measurements}.
  \jt{Boundary-Layer Meteorology}  \bvol{74}~(1),  \pg{19--54}.

\bibitem[Dougherty(1961)]{dougherty1961anisotropy}
{\sc \au{Dougherty, JP}} \yr{1961}  \at{The anisotropy of turbulence at the
  meteor level}.  \jt{Journal of Atmospheric and Terrestrial Physics}
  \bvol{21}~(2-3),  \pg{210--213}.

\bibitem[Fiedler \& Panofsky(1970)]{fiedler1970atmospheric}
{\sc \au{Fiedler, Franz} \& \au{Panofsky, Hans~A}} \yr{1970}  \at{Atmospheric
  scales and spectral gaps}.  \jt{Bulletin of the American Meteorological
  Society}  \bvol{51}~(12),  \pg{1114--1120}.

\bibitem[Flores \& Riley(2011)]{flores2011analysis}
{\sc \au{Flores, O} \& \au{Riley, JJ}} \yr{2011}  \at{Analysis of turbulence
  collapse in the stably stratified surface layer using direct numerical
  simulation}.  \jt{Boundary-Layer Meteorology}  \bvol{139}~(2),
  \pg{241--259}.

\bibitem[Gohari \& Sarkar(2017)]{gohari2017direct}
{\sc \au{Gohari, SM~Iman} \& \au{Sarkar, Sutanu}} \yr{2017}  \at{Direct
  numerical simulation of turbulence collapse and rebirth in stably stratified
  ekman flow}.  \jt{Boundary-Layer Meteorology}  \bvol{162}~(3),
  \pg{401--426}.

\bibitem[Grachev {\em et~al.\/}(2013)Grachev, Andreas, Fairall, Guest \&
  Persson]{grachev2013critical}
{\sc \au{Grachev, Andrey~A}, \au{Andreas, Edgar~L}, \au{Fairall,
  Christopher~W}, \au{Guest, Peter~S} \& \au{Persson, P Ola~G}} \yr{2013}
  \at{The critical richardson number and limits of applicability of local
  similarity theory in the stable boundary layer}.  \jt{Boundary-layer
  meteorology}  \bvol{147}~(1),  \pg{51--82}.

\bibitem[Grachev {\em et~al.\/}(2015)Grachev, Andreas, Fairall, Guest \&
  Persson]{grachev2015similarity}
{\sc \au{Grachev, Andrey~A}, \au{Andreas, Edgar~L}, \au{Fairall,
  Christopher~W}, \au{Guest, Peter~S} \& \au{Persson, P Ola~G}} \yr{2015}
  \at{Similarity theory based on the dougherty--ozmidov length scale}.
  \jt{Quarterly Journal of the Royal Meteorological Society}  \bvol{141}~(690),
   \pg{1845--1856}.

\bibitem[Holt {\em et~al.\/}(1992)Holt, Koseff \& Ferziger]{holt1992numerical}
{\sc \au{Holt, Steven~E}, \au{Koseff, Jeffrey~R} \& \au{Ferziger, Joel~H}}
  \yr{1992}  \at{A numerical study of the evolution and structure of
  homogeneous stably stratified sheared turbulence}.  \jt{Journal of Fluid
  Mechanics}  \bvol{237},  \pg{499--539}.

\bibitem[Jim{\'e}nez \& Cuxart(2005)]{jimenez2005large}
{\sc \au{Jim{\'e}nez, MA} \& \au{Cuxart, J}} \yr{2005}  \at{Large-eddy
  simulations of the stable boundary layer using the standard kolmogorov
  theory: Range of applicability}.  \jt{Boundary-Layer Meteorology}
  \bvol{115}~(2),  \pg{241--261}.

\bibitem[Kaimal(1973)]{kaimal1973turbulenece}
{\sc \au{Kaimal, JCj}} \yr{1973}  \at{Turbulenece spectra, length scales and
  structure parameters in the stable surface layer}.  \jt{Boundary-Layer
  Meteorology}  \bvol{4}~(1-4),  \pg{289--309}.

\bibitem[Kaimal {\em et~al.\/}(1972)Kaimal, Wyngaard, Izumi \&
  Cot{\'e}]{kaimal1972spectral}
{\sc \au{Kaimal, J.~C.}, \au{Wyngaard, J.}, \au{Izumi, Y.} \& \au{Cot{\'e},
  O.R.}} \yr{1972}  \at{Spectral characteristics of surface-layer turbulence}.
  \jt{Quarterly Journal of the Royal Meteorological Society}  \bvol{98}~(417),
  \pg{563--589}.

\bibitem[Katul {\em et~al.\/}(2016)Katul, Li, Liu \&
  Assouline]{katul2016deviations}
{\sc \au{Katul, G.~G.}, \au{Li, D.}, \au{Liu, H.} \& \au{Assouline, S.}}
  \yr{2016}  \at{Deviations from unity of the ratio of the turbulent schmidt to
  prandtl numbers in stratified atmospheric flows over water surfaces}.
  \jt{Physical Review Fluids}  \bvol{1}~(3),  \pg{034401}.

\bibitem[Katul {\em et~al.\/}(2012)Katul, Porporato \&
  Nikora]{katul2012existence}
{\sc \au{Katul, G.~G.}, \au{Porporato, A.} \& \au{Nikora, V.}} \yr{2012}
  \at{Existence of k-1 power-law scaling in the equilibrium regions of
  wall-bounded turbulence explained by heisenberg's eddy viscosity}.
  \jt{Physical Review E}  \bvol{86}~(6),  \pg{066311}.

\bibitem[Katul {\em et~al.\/}(2014)Katul, Porporato, Shah \&
  Bou-Zeid]{katul2014two}
{\sc \au{Katul, Gabriel~G}, \au{Porporato, Amilcare}, \au{Shah, Stimit} \&
  \au{Bou-Zeid, Elie}} \yr{2014}  \at{Two phenomenological constants explain
  similarity laws in stably stratified turbulence}.  \jt{Physical Review E}
  \bvol{89}~(2),  \pg{023007}.

\bibitem[Khani \& Waite(2014)]{khani2014buoyancy}
{\sc \au{Khani, S.} \& \au{Waite, M.~L.}} \yr{2014}  \at{Buoyancy scale effects
  in large-eddy simulations of stratified turbulence}.  \jt{Journal of Fluid
  Mechanics}  \bvol{754},  \pg{75--97}.

\bibitem[Kimura \& Herring(2012)]{kimura2012energy}
{\sc \au{Kimura, Yoshifumi} \& \au{Herring, JR}} \yr{2012}  \at{Energy spectra
  of stably stratified turbulence}.  \jt{Journal of Fluid Mechanics}
  \bvol{698},  \pg{19--50}.

\bibitem[Kolmogorov(1941)]{kolmogorov1941}
{\sc \au{Kolmogorov, Andrey~Nikolaevich}} \yr{1941}  \at{The local structure of
  turbulence in incompressible viscous fluid for very large reynolds numbers}
  \bvol{30}~(4),  \pg{299--303}.

\bibitem[Komori \& Nagata(1996)]{komori1996effects}
{\sc \au{Komori, Satoru} \& \au{Nagata, Kouji}} \yr{1996}  \at{Effects of
  molecular diffusivities on counter-gradient scalar and momentum transfer in
  strongly stable stratification}.  \jt{Journal of Fluid Mechanics}
  \bvol{326},  \pg{205--237}.

\bibitem[Kunkel \& Marusic(2006)]{kunkel2006study}
{\sc \au{Kunkel, Gary~J} \& \au{Marusic, Ivan}} \yr{2006}  \at{Study of the
  near-wall-turbulent region of the high-reynolds-number boundary layer using
  an atmospheric flow}.  \jt{Journal of Fluid Mechanics}  \bvol{548},
  \pg{375--402}.

\bibitem[Li \& Bou-Zeid(2011)]{li2011coherent}
{\sc \au{Li, Dan} \& \au{Bou-Zeid, Elie}} \yr{2011}  \at{Coherent structures
  and the dissimilarity of turbulent transport of momentum and scalars in the
  unstable atmospheric surface layer}.  \jt{Boundary-layer meteorology}
  \bvol{140}~(2),  \pg{243--262}.

\bibitem[Li {\em et~al.\/}(2015)Li, Katul \& Bou-Zeid]{li2015turbulent}
{\sc \au{Li, Dan}, \au{Katul, Gabriel~G} \& \au{Bou-Zeid, Elie}} \yr{2015}
  \at{Turbulent energy spectra and cospectra of momentum and heat fluxes in the
  stable atmospheric surface layer}.  \jt{Boundary-layer meteorology}
  \bvol{157}~(1),  \pg{1--21}.

\bibitem[Li {\em et~al.\/}(2016)Li, Salesky \& Banerjee]{li2016connections}
{\sc \au{Li, Dan}, \au{Salesky, Scott~T} \& \au{Banerjee, Tirtha}} \yr{2016}
  \at{Connections between the ozmidov scale and mean velocity profile in stably
  stratified atmospheric surface layers}.  \jt{Journal of Fluid Mechanics}
  \bvol{797}.

\bibitem[Li {\em et~al.\/}(2018)Li, Bou-Zeid, Vercauteren \&
  Parlange]{li2018signatures}
{\sc \au{Li, Qi}, \au{Bou-Zeid, Elie}, \au{Vercauteren, Nikki} \& \au{Parlange,
  Marc}} \yr{2018}  \at{Signatures of air--wave interactions over a large
  lake}.  \jt{Boundary-Layer Meteorology}  \bvol{167}~(3),  \pg{445--468}.

\bibitem[Lilly(1983)]{lilly1983stratified}
{\sc \au{Lilly, Douglas~K}} \yr{1983}  \at{Stratified turbulence and the
  mesoscale variability of the atmosphere}.  \jt{Journal of the Atmospheric
  Sciences}  \bvol{40}~(3),  \pg{749--761}.

\bibitem[Lindborg(2006)]{lindborg2006energy}
{\sc \au{Lindborg, Erik}} \yr{2006}  \at{The energy cascade in a strongly
  stratified fluid}.  \jt{Journal of Fluid Mechanics}  \bvol{550},
  \pg{207--242}.

\bibitem[Lumley(1964)]{lumley1964spectrum}
{\sc \au{Lumley, JL}} \yr{1964}  \at{The spectrum of nearly inertial turbulence
  in a stably stratified fluid}.  \jt{Journal of the Atmospheric Sciences}
  \bvol{21}~(1),  \pg{99--102}.

\bibitem[Lumley(1965)]{lumley1965theoretical}
{\sc \au{Lumley, John~L}} \yr{1965}  \at{Theoretical aspects of research on
  turbulence in stratified flows}.  \jt{Atmospheric turbulence and radio wave
  propagation. Nauka, Moscow}  \pg{pp. 105--110}.

\bibitem[Lumley \& Panofsky(1964)]{lumley1964structure}
{\sc \au{Lumley, J.~L.} \& \au{Panofsky, H.~A.}} \yr{1964} {\em The structure
  of atmospheric turbulence\/}.  \publ{John Wiley \& Sons}.

\bibitem[Maffioli(2017)]{maffioli2017vertical}
{\sc \au{Maffioli, Andrea}} \yr{2017}  \at{Vertical spectra of stratified
  turbulence at large horizontal scales}.  \jt{Physical Review Fluids}
  \bvol{2}~(10),  \pg{104802}.

\bibitem[Maffioli \& Davidson(2016)]{maffioli2016dynamics}
{\sc \au{Maffioli, A} \& \au{Davidson, PA}} \yr{2016}  \at{Dynamics of
  stratified turbulence decaying from a high buoyancy reynolds number}.
  \jt{Journal of Fluid Mechanics}  \bvol{786},  \pg{210--233}.

\bibitem[Mahrt(1985)]{mahrt1985vertical}
{\sc \au{Mahrt, L}} \yr{1985}  \at{Vertical structure and turbulence in the
  very stable boundary layer}.  \jt{Journal of the Atmospheric Sciences}
  \bvol{42}~(22),  \pg{2333--2349}.

\bibitem[Mahrt(1998)]{mahrt1998stratified}
{\sc \au{Mahrt, Larry}} \yr{1998}  \at{Stratified atmospheric boundary layers
  and breakdown of models}.  \jt{Theoretical and computational fluid dynamics}
  \bvol{11}~(3),  \pg{263--279}.

\bibitem[Mahrt(1999)]{mahrt1999stratified}
{\sc \au{Mahrt, Larry}} \yr{1999}  \at{Stratified atmospheric boundary layers}.
   \jt{Boundary-Layer Meteorology}  \bvol{90}~(3),  \pg{375--396}.

\bibitem[Mahrt(2014)]{mahrt2014stably}
{\sc \au{Mahrt, L}} \yr{2014}  \at{Stably stratified atmospheric boundary
  layers}.  \jt{Annual Review of Fluid Mechanics}  \bvol{46},  \pg{23--45}.

\bibitem[Monin \& Obukhov(1954)]{monin1954basic}
{\sc \au{Monin, AS} \& \au{Obukhov, AMF}} \yr{1954}  \at{Basic laws of
  turbulent mixing in the surface layer of the atmosphere}.  \jt{Contrib.
  Geophys. Inst. Acad. Sci. USSR}  \bvol{151}~(163),  \pg{e187}.

\bibitem[Monin \& Yaglom(1975)]{monin2013statistical}
{\sc \au{Monin, AS} \& \au{Yaglom, AM}} \yr{1975} {\em Statistical fluid
  mechanics, volume II: Mechanics of turbulence\/}, ,  \vol{vol.~2}.  \publ{MIT
  Press}.

\bibitem[Muschinski {\em et~al.\/}(2004)Muschinski, Frehlich \&
  Balsley]{muschinski2004small}
{\sc \au{Muschinski, Andreas}, \au{Frehlich, Rod~G} \& \au{Balsley, Ben~B}}
  \yr{2004}  \at{Small-scale and large-scale intermittency in the nocturnal
  boundary layer and the residual layer}.  \jt{Journal of Fluid Mechanics}
  \bvol{515},  \pg{319--351}.

\bibitem[Nastrom \& Gage(1985)]{nastrom1985climatology}
{\sc \au{Nastrom, GD} \& \au{Gage, K~So}} \yr{1985}  \at{A climatology of
  atmospheric wavenumber spectra of wind and temperature observed by commercial
  aircraft}.  \jt{Journal of the atmospheric sciences}  \bvol{42}~(9),
  \pg{950--960}.

\bibitem[Nieuwstadt(1984)]{nieuwstadt1984turbulent}
{\sc \au{Nieuwstadt, Frans~TM}} \yr{1984}  \at{The turbulent structure of the
  stable, nocturnal boundary layer}.  \jt{Journal of the atmospheric sciences}
  \bvol{41}~(14),  \pg{2202--2216}.

\bibitem[Obukhov(1946)]{obukhov1946turbulence}
{\sc \au{Obukhov, AM}} \yr{1946}  \at{Turbulence in an atmosphere with a
  non-uniform temperature}.  \jt{Trudy Inst. Teoret. Geophys. Akad. Nauk SSSR}
  \bvol{1},  \pg{95--115}.

\bibitem[Ohya {\em et~al.\/}(1997)Ohya, Neff \& Meroney]{ohya1997turbulence}
{\sc \au{Ohya, Yuji}, \au{Neff, David~E} \& \au{Meroney, Robert~N}} \yr{1997}
  \at{Turbulence structure in a stratified boundary layer under stable
  conditions}.  \jt{Boundary-Layer Meteorology}  \bvol{83}~(1),  \pg{139--162}.

\bibitem[Ozmidov(1965)]{ozmidov1965turbulent}
{\sc \au{Ozmidov, RV}} \yr{1965}  \at{On the turbulent exchange in a stably
  stratified ocean. izv. acad. sci. ussr}.  \jt{Atmos. Oceanic Phys.}
  \bvol{1},  \pg{861--871}.

\bibitem[Petenko {\em et~al.\/}((in press))Petenko, Argentini, Casasanta,
  Genthon \& Kallistratova]{Petenko2018stable}
{\sc \au{Petenko, Igor}, \au{Argentini, Stefania}, \au{Casasanta, Giampietro},
  \au{Genthon, Christophe} \& \au{Kallistratova, Margarita}} \yr{(in press)}
  \at{Stable surface-based turbulent layer during the polar winter at dome c,
  antarctica: Sodar and in-situ observations}.  \jt{Boundary-Layer Meteorology}
  .

\bibitem[Phillips(1965)]{phillips1965bolgiano}
{\sc \au{Phillips, OM}} \yr{1965}  \at{On the bolgiano and lumley-shur theories
  of the buoyancy subrange}.  \jt{Atmospheric Turbulence and Radio Wave
  Propagation}  \pg{pp. 121--128}.

\bibitem[Pope(2000)]{pope2000turbulent}
{\sc \au{Pope, SB}} \yr{2000} {\em Turbulent Flows\/}.  \publ{Cambridge
  University Press}.

\bibitem[Reiter \& Burns(1965)]{reiter1965atmospheric}
{\sc \au{Reiter, Elmar~R} \& \au{Burns, Anne}} \yr{1965}  \at{Atmospheric
  structure and clear-air turbulence}.  \jt{Atmospheric science technical
  paper; no. 65} .

\bibitem[Riley \& DeBruynkops(2003)]{riley2003dynamics}
{\sc \au{Riley, James~J} \& \au{DeBruynkops, Stephen~M}} \yr{2003}
  \at{Dynamics of turbulence strongly influenced by buoyancy}.  \jt{Physics of
  Fluids}  \bvol{15}~(7),  \pg{2047--2059}.

\bibitem[Riley \& Lindborg(2008)]{riley2008stratified}
{\sc \au{Riley, James~J} \& \au{Lindborg, Erik}} \yr{2008}  \at{Stratified
  turbulence: A possible interpretation of some geophysical turbulence
  measurements}.  \jt{Journal of the Atmospheric Sciences}  \bvol{65}~(7),
  \pg{2416--2424}.

\bibitem[Schumann(1987)]{schumann1987countergradient}
{\sc \au{Schumann, U}} \yr{1987}  \at{The countergradient heat flux in
  turbulent stratified flows}.  \jt{Nuclear Engineering and Design}
  \bvol{100}~(3),  \pg{255--262}.

\bibitem[Schumann \& Gerz(1995)]{schumann1995turbulent}
{\sc \au{Schumann, Ulrich} \& \au{Gerz, Thomas}} \yr{1995}  \at{Turbulent
  mixing in stably stratified shear flows}.  \jt{Journal of Applied
  Meteorology}  \bvol{34}~(1),  \pg{33--48}.

\bibitem[Selker {\em et~al.\/}(2006)Selker, van~de Giesen, Westhoff, Luxemburg
  \& Parlange]{selker2006fiber}
{\sc \au{Selker, John}, \au{van~de Giesen, Nick}, \au{Westhoff, Martijn},
  \au{Luxemburg, Wim} \& \au{Parlange, Marc~B}} \yr{2006}  \at{Fiber optics
  opens window on stream dynamics}.  \jt{Geophysical Research Letters}
  \bvol{33}~(24).

\bibitem[Shah \& Bou-Zeid(2014)]{shah2014direct}
{\sc \au{Shah, S.~K.} \& \au{Bou-Zeid, E.}} \yr{2014}  \at{Direct numerical
  simulations of turbulent ekman layers with increasing static stability:
  modifications to the bulk structure and second-order statistics}.
  \jt{Journal of Fluid Mechanics}  \bvol{760},  \pg{494--539}.

\bibitem[Smeets {\em et~al.\/}(1998)Smeets, Duynkerke \&
  Vugts]{smeets1998turbulence}
{\sc \au{Smeets, CJPP}, \au{Duynkerke, PG} \& \au{Vugts, HF}} \yr{1998}
  \at{Turbulence characteristics of the stable boundary layer over a
  mid-latitude glacier. part i: A combination of katabatic and large-scale
  forcing}.  \jt{Boundary-Layer Meteorology}  \bvol{87}~(1),  \pg{117--145}.

\bibitem[Smyth \& Moum(2000)]{smyth2000length}
{\sc \au{Smyth, William~D} \& \au{Moum, James~N}} \yr{2000}  \at{Length scales
  of turbulence in stably stratified mixing layers}.  \jt{Physics of Fluids}
  \bvol{12}~(6),  \pg{1327--1342}.

\bibitem[Sreenivasan(1995)]{sreenivasan1995universality}
{\sc \au{Sreenivasan, K.~R.}} \yr{1995}  \at{On the universality of the
  kolmogorov constant}.  \jt{Physics of Fluids}  \bvol{7}~(11),
  \pg{2778--2784}.

\bibitem[Taylor(1938)]{taylor1938spectrum}
{\sc \au{Taylor, G.~I.}} \yr{1938} The spectrum of turbulence.  \bt{In {\em
  Proceedings of the Royal Society of London A: Mathematical, Physical and
  Engineering Sciences\/}}, ,  \vol{vol. 164},  \pg{pp. 476--490}. The Royal
  Society.

\bibitem[Torrence \& Compo(1998)]{torrence1998practical}
{\sc \au{Torrence, Christopher} \& \au{Compo, Gilbert~P}} \yr{1998}  \at{A
  practical guide to wavelet analysis}.  \jt{Bulletin of the American
  Meteorological society}  \bvol{79}~(1),  \pg{61--78}.

\bibitem[Townsend(1958)]{townsend1958turbulent}
{\sc \au{Townsend, AA}} \yr{1958}  \at{Turbulent flow in a stably stratified
  atmosphere}.  \jt{Journal of Fluid Mechanics}  \bvol{3}~(4),  \pg{361--372}.

\bibitem[Townsend(1976)]{townsend1976structure}
{\sc \au{Townsend, Albert~A}} \yr{1976} {\em The structure of turbulent shear
  flow\/}.  \publ{Cambridge university press}.

\bibitem[Tulloch \& Smith(2006)]{tulloch2006theory}
{\sc \au{Tulloch, R} \& \au{Smith, KS}} \yr{2006}  \at{A theory for the
  atmospheric energy spectrum: Depth-limited temperature anomalies at the
  tropopause}.  \jt{Proceedings of the National Academy of Sciences}
  \bvol{103}~(40),  \pg{14690--14694}.

\bibitem[Tyler {\em et~al.\/}(2009)Tyler, Selker, Hausner, Hatch, Torgersen,
  Thodal \& Schladow]{tyler2009environmental}
{\sc \au{Tyler, Scott~W}, \au{Selker, John~S}, \au{Hausner, Mark~B}, \au{Hatch,
  Christine~E}, \au{Torgersen, Thomas}, \au{Thodal, Carl~E} \& \au{Schladow,
  S~Geoffrey}} \yr{2009}  \at{Environmental temperature sensing using raman
  spectra dts fiber-optic methods}.  \jt{Water Resources Research}
  \bvol{45}~(4).

\bibitem[Vercauteren {\em et~al.\/}(2008)Vercauteren, Bou-Zeid, Parlange,
  Lemmin, Huwald, Selker \& Meneveau]{vercauteren2008subgrid}
{\sc \au{Vercauteren, N.}, \au{Bou-Zeid, E.}, \au{Parlange, M.~B.}, \au{Lemmin,
  U.}, \au{Huwald, H.}, \au{Selker, J.} \& \au{Meneveau, C.}} \yr{2008}
  \at{Subgrid-scale dynamics of water vapour, heat, and momentum over a lake}.
  \jt{Boundary-layer meteorology}  \bvol{128}~(2),  \pg{205--228}.

\bibitem[Vignon {\em et~al.\/}(2017{\natexlab{{\em a\/}}})Vignon, Genthon,
  Barral, Amory, Picard, Gall{\'e}e, Casasanta \&
  Argentini]{vignon2017momentum}
{\sc \au{Vignon, E.}, \au{Genthon, C.}, \au{Barral, H.}, \au{Amory, C.},
  \au{Picard, G.}, \au{Gall{\'e}e, H.}, \au{Casasanta, G.} \& \au{Argentini,
  S.}} \yr{2017{\natexlab{{\em a\/}}}}  \at{Momentum-and heat-flux
  parametrization at dome c, antarctica: A sensitivity study}.
  \jt{Boundary-Layer Meteorology}  \bvol{162}~(2),  \pg{341--367}.

\bibitem[Vignon {\em et~al.\/}(2017{\natexlab{{\em b\/}}})Vignon, van~de Wiel,
  van Hooijdonk, Genthon, van~der Linden, van Hooft, Baas, Maurel, Traull{\'e}
  \& Casasanta]{vignon2017stable}
{\sc \au{Vignon, Etienne}, \au{van~de Wiel, Bas~JH}, \au{van Hooijdonk,
  Ivo~GS}, \au{Genthon, Christophe}, \au{van~der Linden, Steven~JA}, \au{van
  Hooft, J~Antoon}, \au{Baas, Peter}, \au{Maurel, William}, \au{Traull{\'e},
  Olivier} \& \au{Casasanta, Giampietro}} \yr{2017{\natexlab{{\em b\/}}}}
  \at{Stable boundary-layer regimes at dome c, antarctica: observation and
  analysis}.  \jt{Quarterly Journal of the Royal Meteorological Society}
  \bvol{143}~(704),  \pg{1241--1253}.

\bibitem[Waite(2011)]{waite2011stratified}
{\sc \au{Waite, M.~L.}} \yr{2011}  \at{Stratified turbulence at the buoyancy
  scale}.  \jt{Physics of Fluids}  \bvol{23}~(6),  \pg{066602}.

\bibitem[Waite(2014)]{waite2014direct}
{\sc \au{Waite, M.~L.}} \yr{2014}  \at{Direct numerical simulations of
  laboratory-scale stratified turbulence}.  \jt{Modelling Atmospheric and
  Oceanic Flows: Insights from Laboratory Experiments (ed. T. von Larcher \& P.
  Williams), American Geophysical Union} .

\bibitem[Waite \& Bartello(2004)]{waite2004stratified}
{\sc \au{Waite, M.~L.} \& \au{Bartello, P.}} \yr{2004}  \at{Stratified
  turbulence dominated by vortical motion}.  \jt{Journal of Fluid Mechanics}
  \bvol{517},  \pg{281--308}.

\bibitem[Waite \& Bartello(2006)]{waite2006stratified}
{\sc \au{Waite, Michael~L} \& \au{Bartello, Peter}} \yr{2006}  \at{Stratified
  turbulence generated by internal gravity waves}.  \jt{Journal of fluid
  Mechanics}  \bvol{546},  \pg{313--339}.

\bibitem[Weinstock(1978)]{weinstock1978theory}
{\sc \au{Weinstock, J}} \yr{1978}  \at{On the theory of turbulence in the
  buoyancy subrange of stably stratified flows}.  \jt{Journal of the
  atmospheric sciences}  \bvol{35}~(4),  \pg{634--649}.

\bibitem[Weinstock(1985)]{weinstock1985theory}
{\sc \au{Weinstock, J}} \yr{1985}  \at{On the theory of temperature spectra in
  a stably stratified fluid}.  \jt{Journal of physical oceanography}
  \bvol{15}~(4),  \pg{475--477}.

\bibitem[Zilitinkevich {\em et~al.\/}(2007)Zilitinkevich, Elperin, Kleeorin \&
  Rogachevskii]{zilitinkevich2007energy}
{\sc \au{Zilitinkevich, SS}, \au{Elperin, T}, \au{Kleeorin, N} \&
  \au{Rogachevskii, I}} \yr{2007}  \at{Energy-and flux-budget (efb) turbulence
  closure model for stably stratified flows. part i: steady-state, homogeneous
  regimes}.  \bt{In {\em Atmospheric Boundary Layers\/}},  \pg{pp. 11--35}.
  \publ{Springer}.

\bibitem[Zilitinkevich {\em et~al.\/}(2013)Zilitinkevich, Elperin, Kleeorin,
  Rogachevskii \& Esau]{zilitinkevich2013hierarchy}
{\sc \au{Zilitinkevich, SS}, \au{Elperin, T}, \au{Kleeorin, N},
  \au{Rogachevskii, I} \& \au{Esau, I}} \yr{2013}  \at{A hierarchy of
  energy-and flux-budget (efb) turbulence closure models for stably-stratified
  geophysical flows}.  \jt{Boundary-layer meteorology}  \pg{pp. 1--33}.

\end{thebibliography}

\end{document}